\documentclass[%
 aip,
 amsmath,amssymb,
 reprint,%
]{revtex4-1}

\usepackage{graphicx}% Include figure files
\usepackage{dcolumn}% Align table columns on decimal point
\usepackage{bm}% bold math
\usepackage[utf8]{inputenc}
\usepackage[T1]{fontenc}
\usepackage{mathptmx}
\usepackage{etoolbox}
\usepackage{siunitx}

\makeatletter
\def\@email#1#2{%
 \endgroup
 \patchcmd{\titleblock@produce}
  {\frontmatter@RRAPformat}
  {\frontmatter@RRAPformat{\produce@RRAP{*#1\href{mailto:#2}{#2}}}\frontmatter@RRAPformat}
  {}{}
}%
\makeatother
\begin{document}

\title[]{Effects of oxidation and impurities in lithium surfaces on the emitting wall plasma sheath}
\author{Kolter Bradshaw}
\altaffiliation[Electronic mail: ]{kb8689@princeton.edu}
\affiliation{ 
Princeton University Department of Astrophysical Sciences, Princeton, NJ 08544, USA
}%

\author{Ammar Hakim}%
\affiliation{ 
Princeton Plasma Physics Laboratory, Princeton, NJ 08543, USA
}%
\author{Bhuvana Srinivasan}%
 \altaffiliation[Author to whom correspondence should be addressed: ]{srinbhu@uw.edu}
\affiliation{ 
University of Washington, Seattle, WA 98195, USA
}%

\date{\today}

\begin{abstract}
Use of lithium as a surface coating in fusion devices improves plasma performance, but the change in wall properties affects the secondary electron emission properties of the material. Lithium oxidizes easily, which drives the emission yield well above unity. We present here simulations demonstrating the change in sheath structure from monotonic to the nonmonotonic space-charge limited sheath using an energy-dependent data-driven emission model which self-consistently captures both secondary emission and backscattering populations. Increased secondary electron emission from the material has ramifications for the degradation and erosion of the wall. Results shows that the oxidation leads to an increased electron energy flux into the wall, and a reduced ion energy flux. The net transfer of energy to the surface is significantly greater for the oxidized case than for the pure lithium case. High backscattering rates of low-energy particles leads to a high re-emission rate at the wall.
\end{abstract}

\maketitle

\section{Introduction}

In plasma devices, such as the tokamak\cite{artsimovich1972} and Hall thrusters,\cite{goebel2023} where plasma is constrained to flow within a channel, the interaction between the confined plasma and the material surface of the device plays a decisive role in both plasma performance and the durability of the device. High particle fluxes to the wall can lead to high rates of material sputtering and fuel recycling, which degrades plasma performance,\cite{stangeby2000plasma} while erosion of wall material degrades durability and increases the potential of catastrophic device failure.\cite{pitts2005material} The plasma sheath\cite{robertson2013} plays a key role in shielding the wall from high electron fluxes, though it accelerates ions and increases recycling rates. Emission of electrons from the surface into the sheath reduces this shielding by increasing the wall potential,\cite{hobbs1967} allowing higher electron fluxes into the wall. Traditionally, common material choices for the tokamak wall and plasma-facing components are tungsten,\cite{jet1990results,neu2003tungsten,coenen2011melt,kuang2020divertor} graphite,\cite{jet1990results,menard2012overview,kuang2020divertor} and beryllium,\cite{jet1990results} Application of a liquid metal film, such as lithium, has seen increased use in recent years\cite{majeski2009,majeski2010,kugel2012,de2021} due to the high binding rate of lithium with hydrogen, reducing fuel recycling and increasing plasma performance.\cite{zakharov1997} However, while lithium has a low emission rate when pure, it oxidizes easily and begins to emit at much higher rates.\cite{bruining1938,capece2016} The level of exposure required for this oxidation to occur is commonly experienced in fusion devices,\cite{skinner2013plasma} making it an important consideration for any plasma-surface interactions. Understanding the ramifications this may have on wall durability requires a thorough understanding of how the plasma sheath responds to this strong emission. The magnitude of several mechanisms which drive loss of wall fidelity are determined by the sheath. Sputtering of wall material is caused by the impact of high-energy ions.\cite{stangeby2000plasma} High heat fluxes lead to destruction of wall material through macroscopic erosion and melting,\cite{wurz2002macroscopic}.\par
The classical sheath maintains a flux balance at the material surface by accelerating ions and retarding the more mobile electrons. The presence of electron emission adds an outbound flux at the wall, altering the flux balance kept by the sheath,
\begin{equation}
    \Gamma_{pe} = \Gamma_{se} + \Gamma_{i}.
    \label{eq:c5_flux_balance}
\end{equation}
Here, the incoming electron flux from the bulk plasma ($\Gamma_{pe}$) is balanced by the total ion particle flux ($\Gamma_i$) and the outgoing emitted electron flux away from the wall ($\Gamma_{se}$). We define the parameter $\gamma = \Gamma_{se}/\Gamma_{pe}$, to denote the \textit{secondary electron emission coefficient}, or the ratio of emitted flux at the wall to incoming flux from the bulk plasma.\par
The addition of this outbound flux leads to a drop in wall potential as less of a barrier is required to balance Eq.~\ref{eq:c5_flux_balance}. Hobbs \& Wesson\cite{hobbs1967} describe this decay in the sheath with increasing $\Gamma_{se}$, culminating with the flux ratio reaching some critical value $\gamma_c$, where the monotonic sheath is no longer capable of balancing the fluxes and a barrier must form close to the wall to reflect back some portion of $\Gamma_{se}$, denoted as $\Gamma_{ref}$. Noting that by definition $\Gamma_{se} = \gamma\Gamma_{pe}$, the resulting flux balance becomes
\begin{equation}
    \Gamma_{pe}(1 - \gamma) + \Gamma_{ref} = \Gamma_{i}.
    \label{eq:c5_scl_flux_balance}
\end{equation}
This indicates the formation of the so-called space-charge limited (SCL) sheath, with a non-monotonic potential dip near the wall.\cite{hobbs1967} If $\gamma$ grows to exceed unity, the $\Gamma_{pe}$ term becomes negative. Here, two solutions are possible. As before, an SCL is capable of providing sufficient $\Gamma_{ref}$ to maintain the flux balance. However, it is also possible for the polarity of the sheath to reverse entirely with the formation of an inverse sheath, where the wall potential is positive relative to the sheath entrance, electrons are accelerated, and ions are decelerated.\citep{campanell2013} The work done by Campanell \& Umansky\cite{campanell2016,campanell2017} demonstrates that the inverse sheath mode is preferred for all $\gamma > 1$ in the presence of ion-neutral collisional effects, which lead to an accumulation of cold ions in the potential dip of the SCL and drive the transition. Without some source of cold ions, the sheath will remain in the SCL state for all $\gamma > \gamma_c$.\par
Building on past sheath work,\cite{cagas2017_sheath,skolar2023,bradshaw2024} this paper explores the effect of lithium oxidation on the plasma sheath. The purpose is to examine key properties of the emitting sheath with an eye to how they may impact durability of the wall material. Section~\ref{sec:model} discusses the numerical model and treatment of secondary electron emission. Particular attention is given to the estimation of emission behavior in the low energy regime from secondary electron yield data. Simulations are set up in Section~\ref{sec:sims} giving comparisons of clean and oxidized lithium parameters. Section~\ref{sec:results} gives a discussion of the results, with an emphasis on how the different emission properties modify the overall sheath structure and energy fluxes into the wall.\par

\section{Model} \label{sec:model}

\subsection{Discrete Kinetic Equations}

The simulation work shown in this paper is produced by the \texttt{Gkeyll} software,\cite{gkyldocs} using the continuum kinetic approach of applying the discontinuous Galerkin (DG) numerical method\cite{cockburn2001} to evolve the Vlasov-Poisson system of equations,\cite{juno2018discontinuous,hakim2020alias} consisting of the electrostatic Boltzmann equation
\begin{equation}
    \frac{\partial f_s}{\partial t} + \mathbf{v}\cdot\nabla_{\mathbf{x}}f_s + \frac{q_s}{m_s}\mathbf{E}\cdot\nabla_{\mathbf{v}}f_s = \bigg(\frac{\partial f}{\partial t}\bigg)_c + S_{src},\label{eq:vlasov}
\end{equation}
coupled with Poisson's equation through the electric field.\par
The collision term is a Bhatnagar-Gross-Krook (BGK) operator, used to thermalize the presheath plasma. It takes the form
\begin{equation}
    \bigg(\frac{\partial f_s}{\partial t}\bigg)_c = \nu_s\bigg(\frac{n}{n_0}f_{0,s} - f_s\bigg),
\end{equation}
where $n_0$ is the initial density and $\nu_s$ is the collision frequency. This relaxes the distribution towards $f_{0,s}$, the Maxwellian distribution at the initial temperature $T_0$. Temperature is fixed for the collision operator as otherwise losses of energetic electrons to the wall will cause a steady cooling over time, preventing a steady state presheath temperature from developing. Previous work has studied the effects of collisions on the Bohm criterion.\cite{li2022bohm,li2022transport} To prevent the thermalization from extending into the sheath region, which is physically close to collisionless, the collision frequency follows a spatial profile
\begin{equation}
    \nu_s(x) = \frac{\nu_{0,s}}{1 + \exp\bigg(\frac{|x|}{3\lambda_D} - \frac{16}{3}\bigg)},
\end{equation}
which falls off quickly towards zero at around $\approx 40\lambda_D$ from the presheath domain boundary, with $\lambda_D = \sqrt{\frac{\epsilon_0T_0}{n_0q_0^2}}$ being the Debye length.\par
A source term $S_{src}$ is also used to preserve the particle balance,
\begin{equation}
    S_{src} = \frac{\Gamma^+_i}{L_{src}}\bigg(\frac{2(L_{src} - |x|)}{L_{src}}\bigg)\frac{n}{n_0}f_{0,s}.
\end{equation}
For both species, the ion particle flux into the wall $\Gamma^+_i$ is used to preserve quasineutrality. $L_{src} = 40\lambda_D$ is the source length, over which there is a linear decreasing profile. This source region, which includes both the area of particle introduction and high collisions, exists to artificially create a region from which thermal particles will flow into the presheath and replenish particle losses to the wall. The addition of particles in this region will create a non-physical "source sheath"\cite{Procassini_1990} at the presheath edge.

\subsection{Emission Yield Fits to Material Data}

There are two primary modes of electron emission which are expected to occur at the material surface.\cite{jensen2018} Some electrons which impact the surface either lack the energy to penetrate and are reflected elastically, or penetrate but are re-emitted after several scattering events internal to the material while losing some energy. These are the ``backscattered" electrons. The true ``secondary" electrons are emitted directly from the material in response to electron impact. The total ratio of impacting to emitted particles is defined here as the \textit{secondary electron yield} $\delta$. This is distinguished from the SEE coefficient $\gamma$ by being the precise ratio of emitted electrons to impacting electrons at the wall $\delta = \Gamma_{se}/(\Gamma_{ref} + \Gamma_{pe})$, while $\gamma$ neglects the contribution of the reflected electron population and is more characteristic of the flux ratio using the incoming flux measured at the potential minimum. Note that these factors are not often clearly distinguished from each other in the literature as they are here. $\gamma$ is used prominently by sheath theory papers (where it is common to assume only the flux from the bulk plasma contributes to the emission) to characterize the sheath mode which develops, while $\delta$ is a more accurate metric for analyzing the material emission model being used, which does not discriminate between impacting particles based on origin. For a classical sheath where there is no potential barrier reflecting emitted electrons back to the wall, $\gamma$ and $\delta$ are equivalent. Past papers which account for the re-emission of reflected particles have found it useful to split $\gamma$ and $\Gamma_{se}$ and distinguish between the emission driven by ``primary" particles entering the sheath from the bulk plasma and the emission driven by emitted particles reflected back to the wall by the potential barrier.\cite{taccogna2014dust,campanell2015self} Continuum simulations such as those in this work do not lend themselves to this more precise diagnostic as they cannot track the origin of individual particles in the same way as PIC codes.\par
\begin{figure}
    \includegraphics[width=1.0\linewidth]{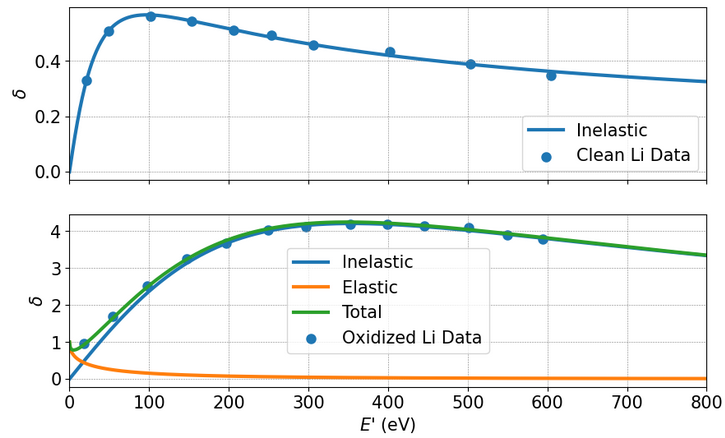}
    \caption{\label{fig:lithium_fit} Yield data from Bruining \& De Boer\cite{bruining1938} for clean (top) and impure (bottom) lithium, with fits to the elastic and inelastic regions of the yield. A fit to the peak of the impure lithium curve results in an increasing undershoot of the emission at low energies, suggesting a rising elastic backscattering coefficient at low energy. This trait is not evident in the sparse clean lithium data, suggesting negligible backscattering at low energy. Figure taken from a Ph.D. dissertation.\cite{bradshaw2024_diss}}
\end{figure}

Data from Bruining \& De Boer\cite{bruining1938} (reproduced in Fig.~\ref{fig:lithium_fit}) shows results for both ``pure" and ``impure" lithium values of $\delta$, with the measured yields being significantly higher in the impure case. Thorough measurements of the electron yield for different oxidized samples of lithium from Capece et al\cite{capece2016} reinforce the choice of the Bruining \& De Boer data as a good upper limit for the emission in cases of high lithium impurity.\par
In order to rigorously fit secondary electron emission models to this data, both backscattering and secondary emission processes must be considered. Ideally, the overall yield curve would be decomposed based on the emission spectrum, where clear distinctions can generally be made on which particles were elastically reflected on impact, and which are ``true" secondary electrons. This data is not present in the literature for lithium, so informed estimates must be made based solely on the yield curve. Noting that elastic backscattering is generally negligible at high energy and rises at low energy,\cite{furman2002} the simplest way of separating the two contributions is to fit the yield curve to the high energy data, and watch for an undershoot at low energy where the backscattering is not being accounted for, fitting to that undershoot for the elastically reflected particles.\par
There is still some ambiguity in the literature around what low energy emission behavior should look like. The work of Cimino et al\cite{cimino2004,larciprete2013} suggests that it is common to see a rise in the emission yield towards unity as energy goes to zero. The accuracy of this has been questioned in light of past experiments,\cite{andronov2013} but in a follow-up paper\cite{cimino2015} on measurements of the yield for copper, Cimino et al observed that the rise appears to be physical for the impure, ``as-received" material samples, while the clean samples are more consistent with previous literature\cite{bronshtein1958, khan1963, yakubova1970} showing a decline towards zero of the yield at low energy.\par
The fitting model for the true secondary electron yield $\delta_{ts}$ comes from Furman \& Pivi,\cite{furman2002}
\begin{equation}
    \delta_{ts}(E', \mu') = \hat{\delta}(\mu')D(E'/\hat{E}(\mu')),
    \label{eq:c4_see_yield}
\end{equation}
\begin{equation}
	\hat{\delta}(\mu') = \hat{\delta}_{ts}[1 + t_1(1 - \mu'^{t_2})],
\end{equation}
\begin{equation}
	\hat{E} = \hat{E}_{ts}[1 + t_3(1 - \mu'^{t_4})],
\end{equation}
\begin{equation}
	D(x) = \frac{sx}{s - 1 + x^s},
\end{equation}
where $\hat{E}_{ts}$, $\hat{\delta}_{ts}$, $t_1$, $t_2$, $t_3$, $t_4$ and $s$ are fitting parameters, with $\hat{E}_{ts}$ and $\hat{\delta}_{ts}$ corresponding to the energy of maximum yield and maximum yield, respectively. The fit is done specifically to the high energy peak region of the data, where the bulk of the emission should be from secondary electron emission.\par
The true secondary emission yield fits to the lithium data in Fig.~\ref{fig:lithium_fit} suggest similar behavior as to that observed by Cimino et al for copper.\cite{cimino2015} The fit to the peak of the impure lithium measurements exhibits a distinct undershoot of the yield that increases with decreasing energy despite excellent correlation at high energy, suggesting a rise in the yield due to backscattering at low energy. Conversely, while the data is too sparse at low energies for full confidence, a similar fit suggests no such upward trend occurs in the clean lithium data.\par
To account for the apparent increase in yield at low energy in the oxidized data, a second fit is done to low energy using the model developed by Cazaux\cite{cazaux2012} to estimate the backscattering yield $\delta_r$,
\begin{equation}
    \begin{split}
    &E_s = E' + E_f + \varphi_r\\
    &G = 1 + \frac{E_s - E'}{E'\mu'^2},\\
    &\delta_r(E') = \frac{(1 - \sqrt{G})^2}{(1 + \sqrt{G})^2},
    \end{split}
\end{equation}
where $E_f$ and $\varphi_r$ are fitting parameters identified with the Fermi energy and work function of the material, respectively. No second fit is required for the clean data as the initial fit matched the data well across the entire energy range. The resulting parameters for both the elastic and inelastic fits are shown in Table~\ref{tb:lithium_fit}\par
Inelastic backscattering is typically approximately constant at high energies,\cite{furman2002} and is therefore folded into the secondary emission instead of being represented separately. As backscattering is directly related to the atomic number $Z$, it is estimated to be quite low at high energy in any event for a low $Z$ material like lithium.\cite{everhart1960}\par
\begin{figure}
    \includegraphics[width=1.0\linewidth]{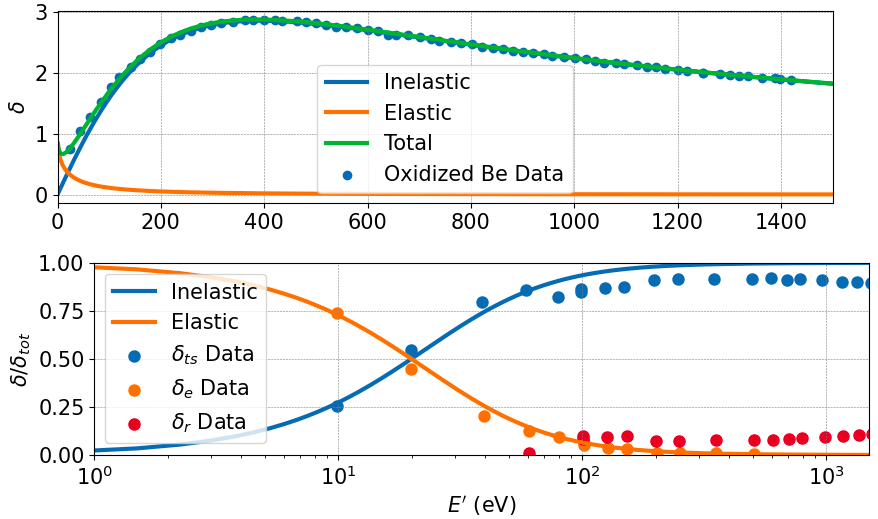}
    \caption{\label{fig:beryllium_fit} Yield data (top) from Shih et al\cite{shih2002secondary} for oxidized beryllium, with fits to the elastic and inelastic regions of the yield matching the procedure done for lithium. Comparison of these fits to the inelastic and elastic region with data from Shih et al giving the decomposition of the yield into the individual emission components\cite{shih2002secondary} validates the fits (bottom). Note that the rediffused electron yield $\delta_r$ is included in the inelastic fit with the true secondary yield $\delta_{ts}$.}
\end{figure}

This fitting approach can be validated by application to a material where experimental measurements have been done decomposing the yield into the component contributions from different emission mechanisms. One such material is beryllium, which is a particularly advantageous choice as it is a prominent tokamak wall material\cite{jet1990results} and exhibits a drastic increase in yield when oxidized,\cite{bruining1938,shih2002secondary} similar to lithium. Fig.~\ref{fig:beryllium_fit} applies the same fitting process detailed above for lithium to oxidized beryllium data from Shih et al\cite{shih2002secondary} The resulting fit preserves the correct proportion $\delta/\delta_{tot}$ for the inelastic (true secondary and inelastic backscattering) vs elastic emission mechanisms.\par
For purposes of comparison to other common tokamak materials, fitting parameters are also given in Table~\ref{tb:lithium_fit} for tungsten\cite{coomes1939total} and graphite.\cite{cazaux2005secondary} No complete data for oxidation effects on the yield of these two materials is present in the literature, though for tungsten experiments suggest the yield significantly \textit{lowers} when oxidized,\cite{vida2003characterization,de2023cold} a reversal of the behavior seen in the other metals considered here. Note that the maximum yields ($\hat{\delta}_{ts}$) of these two materials peak at above unity, but only at sufficiently high energy ($\hat{E}_{ts}$) that it is doubtful this regime will be present once the sheath forms and electrons are significantly decelerated from the initial condition. Note also that, similarly to clean lithium, neither of these two datasets captured any noticeable undershoot at low energy to suggest elastic emission occurring, hence the lack of elastic parameters.

\begin{table}
\centering
\begin{tabular}{ |c||c|c|c|c|c|c|c||c|c|  }
\hline
Material & $\hat{E}_{ts}$ & $\hat{\delta}_{ts}$ & $t_1$ & $t_2$ & $t_3$ & $t_4$ & $s$ & $E_f$ & $\varphi_r$ \\
\hline
Clean Li & $97.18$ & $0.567$ & $0.66$ & $0.8$ & $0.7$ & $1.0$ & $1.42$ & $0$ & $0$ \\
Oxidized Li & $354.52$ & $4.208$ & $0.66$ & $0.8$ & $0.7$ & $1.0$ & $1.79$ & $290.31$ & $144.49$ \\
\hline
Tungsten & $466.06$ & $1.29$ & $0.66$ & $0.8$ & $0.7$ & $1.0$ & $1.72$ & $0$ & $0$ \\
Graphite & $292.17$ & $1.20$ & $0.66$ & $0.8$ & $0.7$ & $1.0$ & $1.67$ & $0$ & $0$ \\
Oxidized Be & $391.97$ & $2.86$ & $0.66$ & $0.8$ & $0.7$ & $1.0$ & $1.66$ & $223.5$ & $76.43$ \\
\hline
\end{tabular}
\caption{Fitting parameters for oxidized and clean lithium boundary condition, as well as three common alternative tokamak materials. The clean case has negligible elastic emission, so those parameters are set to zero.}
\label{tb:lithium_fit}
\end{table}

\subsection{Boundary Condition}

The emission is implemented as a boundary condition where the inelastic and elastic populations are fully modeled.\cite{bradshaw2024} The boundary condition sets the distribution function in a layer of ``ghost" cells outside the domain edge, which are used in the numerical update of the interior ``skin" cells neighboring the material wall. Taking $+$ to denote the incoming direction into the wall, and $-$ the outgoing direction away from the wall, the result is the ghost cell distribution being set to
\begin{equation}
    f_e^{g-}(\mathbf{x}, \mathbf{v}) = f_r(\mathbf{x}, \mathbf{v}) + C_{ts}f_{ts}(\mathbf{v}),
\end{equation}
where $f_r$ is the distribution of elastically reflected particles, $f_{ts}$ is the distribution of emitted secondary electrons, and $C_{ts}$ is the normalization factor which sets the correct flux ratio between the emitted and impacting electrons. The elastic distribution is found by simply scaling the incoming distribution in the skin cell by the elastic yield, 
\begin{equation}
    f_r(\mathbf{x}, \mathbf{v}) = \delta_r(\mathbf{v})f_e^{s+}(\mathbf{x}^-, \mathbf{v}^-),
\end{equation}
where the coordinates $\mathbf{x}^-$, $\mathbf{v}^-$ indicate a transformation from incoming to outgoing coordinates (i.e., $\mathbf{x}^-, \mathbf{v}^- = (-x, y, z), (-v_x, v_y, v_z)$ at the $x$-boundary). Since there is no complete data available for the energetic distribution of secondary electrons for lithium, it is best estimated using the theory from Chung \& Everhart\cite{chung1974}
\begin{equation}
    f_{ts}(\mathbf{v}) = \frac{E(\mathbf{v})}{(E(\mathbf{v}) + \varphi)^4},
\end{equation}
where $\varphi$, the material work function, is the sole material parameter dependence. For lithium, this value is $\varphi\approx\SI{3}{\electronvolt}$.\cite{haynes2016} The Chung-Everhart model places the energy at which the peak of the emitted distribution is located at $\varphi/3 = \SI{1}{\electronvolt}$.
For the Chung-Everhart spectrum, the normalization factor is calculated to be
\begin{equation}
    C_{ts} = \frac{6m_e}{q_0}\varphi^2\bar{\delta}_{ts}\Gamma_e^+,
\end{equation}
where $\bar{\delta}_{ts}$ is the weighted average of the secondary electron yield over the incoming velocity space,
\begin{equation}
    \bar{\delta}_{ts} = \frac{\sum_{j=1}^{N^+_v}\Gamma_e^j\delta_{ts}(v^j_c)}{\Gamma_e^+}.
\end{equation}
Here, $\Gamma_e^j$ is the electron flux of the distribution function into the boundary cell $j$, and $v^j_c$ is the cell center velocity.\par

\section{Simulations} \label{sec:sims}

The choice of 1X1V domain is $N_x = 960$ cells in configuration space, with bounds of $[0,128\lambda_D]$. In velocity space, there are $N_v = 512$ cells with bounds of $[-4v_{t,e}, 4v_{t,e}]$ for the electrons, and $[-6v_{t,i}, 6v_{t,i}]$ for the ions, where $v_{t,s} = \sqrt{\frac{T_0}{m_s}}$ is the thermal velocity of the particle species. The plasma is initialized to a uniform temperature of $T_0 = \SI{150}{\electronvolt}$ for both electrons and ions. The initial density is $n_0 = \SI{1.0e16}{\metre^{-3}}$. The source length is $L_s = 40\lambda_D$. The temperature and density are estimates from gyrokinetic simulations\cite{shukla2023gk} of the Lithium Tokamak Experiment\cite{majeski2009} (LTX) scrape-off layer. In addition to being a relevant application, this choice of parameter regime has the advantage of being low enough energy that the resolving the emitted electron distribution function in velocity space with the chosen number of cells is not an issue, as it often is for temperatures exceeding $\SI{1}{\kilo\electronvolt}$.\cite{bradshaw2024}\par
Four different lithium cases are presented here, distinct by choice of material fit and collisionality. Case 1 uses the clean lithium boundary parameters, and a high collisionality of $\tilde{\nu}_{0,s} = \nu_{0,s}\lambda_D/v_{t,s} = 0.2$. Case 2 uses the same high collisionality with oxidized lithium parameters. Case 3 is oxidized, with the marginally collisional $\tilde{\nu}_{0,s} = 0.02$. This case is included to demonstrate possible behavior when turbulence propagates into the sheath region, whether as a result of turbulent activity in the core plasma or due to emitted beams propagating from the other wall such as might occur in a narrow channel application (such as the Hall thruster). Case 4 is the same as case 2, but with the elastic emission parameters set to zero. This final case is included to highlight how the inclusion of low-energy emission emission physics alters the sheath behavior. In addition to the lithium cases, simulations were run for tungsten ($\varphi\approx4.5$\cite{davisson1922thermionic}), graphite ($\varphi\approx4.5$\cite{rut2020graphene}), and oxidized beryllium ($\varphi\approx3.8$\cite{jerner1970effect}). Detailed plots are not presented for these cases, but important results are tabulated in Table~\ref{tb:simulation_results} for comparison.\par
If emission is applied at full strength immediately to the uniform plasma, it triggers the sudden introduction of a sharp density gradient within the domain edge cell due to the large number of energetic particles being allowed to reach the wall and emit. This can cause serious positivity issues, so for these simulations the boundary condition is set to increase from zero emission at $t\omega_{pe} = 0$ to full emission at $t_b\omega_{pe} = 1000$. This is done by scaling the normalization factor for the emitted distribution by a factor of $\sin{(\frac{\pi t}{2t_b})}$ for $t < t_b$. This allows the plasma to relax from the initial uniform condition to the sheath where fewer high energy particles are present without compromising the stability of the simulation.\par
There is no mechanism present in the simulations to provide cold ions in the sheath region to drive a potential transition to an inverse sheath. Neutral effects are currently not implemented extensively in the Vlasov-Poisson code being used, and attempts to replicate the effects of similar processes on the ions using a non-physical extension of collisions or the source term into the sheath lead to inconclusive results. Accumulation of trapped ions and increases in sheath potentials are observed, but none of these tests result in an inverse sheath transition on simulation time scales. Due to a lack of confidence in any conclusions on the expected long-term behavior of the sheath drawn from these simulations, they are omitted from this work.\par
Trapped ions can appear organically in the potential dip if the time relaxation is not used and full emission is permitted immediately. This occurs during the initial relaxation from the uniform state, where the potential dip forms rapidly enough to capture some portion of the cold ions before the ion distribution fully accelerates to higher speeds. In one previous simulation, this initial trapping is even seemingly enough to drive a reverse sheath transition,\cite{bradshaw2024_diss} though this only occurs in a single known case where the presheath temperature is not fixed and significant cooling across the simulation domain also occurs. In any case, this trapped population remains mostly static as no additional source is present, and does not represent a physically significant result.\par

\begin{figure}
    \includegraphics[width=1.0\linewidth]{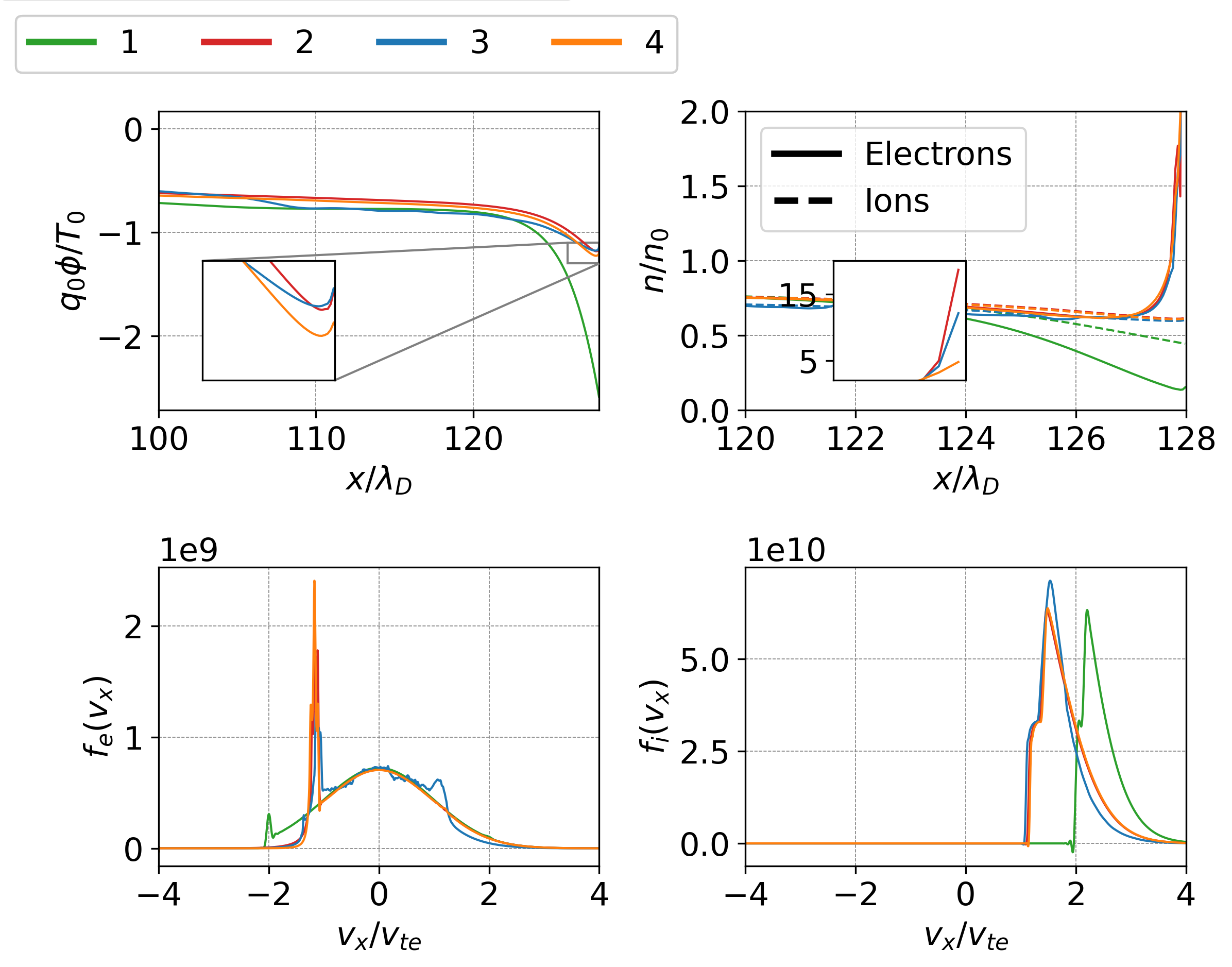}
    \caption{\label{fig:sheath_profiles} Profiles for the potential, density, and species distribution functions  at $t\omega_{pe} = 8000$ for clean lithium with strong collisions (1), oxidized lithium with strong collisions (2), oxidized lithium with weak collisions (3), and oxidized lithium with the elastic emission removed (4). The electron distributions are averaged over the presheath from $60\lambda_D$ to $100\lambda_D$, while the ion distributions are taken at the wall boundary. The oxidized lithium cases form SCL sheaths, with high electron accumulation at the wall and higher intensity emitted beams are visible in the distribution. The marginally collisional case is noticeably less Maxwellian in the electron distribution, and the turbulence leads to a flattening of the center of the distribution, and a beam propagating back from the left boundary is visible. The clean case, which forms a classical sheath, accelerates ions more than the SCL cases due to the greater sheath potential drop.}
\end{figure}

\begin{figure}
    \includegraphics[width=1.0\linewidth]{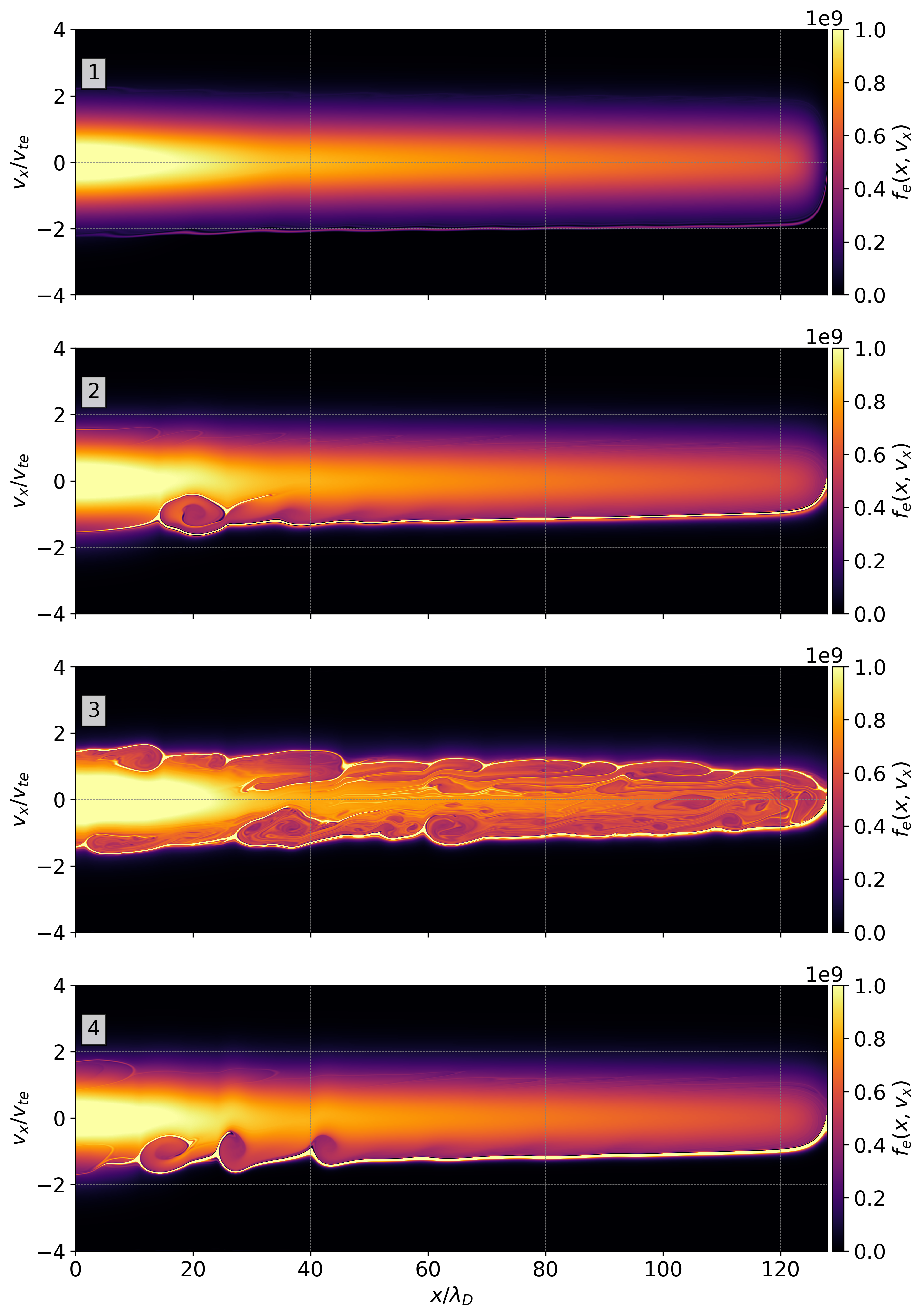}
    \caption{\label{fig:electron_dist} Electron distribution function at $t\omega_{pe} = 8000$ for clean lithium with strong collisions (1), oxidized lithium with strong collisions (2), oxidized lithium with weak collisions (3), and oxidized lithium with the elastic emission removed (4). Emission drives a beam of electrons from the wall which penetrate the presheath, exciting two-stream effects in cases of strong emission. The weakly collisional oxidized case experiences strong turbulence when these modes propagate back to the sheath.}
\end{figure}

\begin{figure}
    \includegraphics[width=1.0\linewidth]{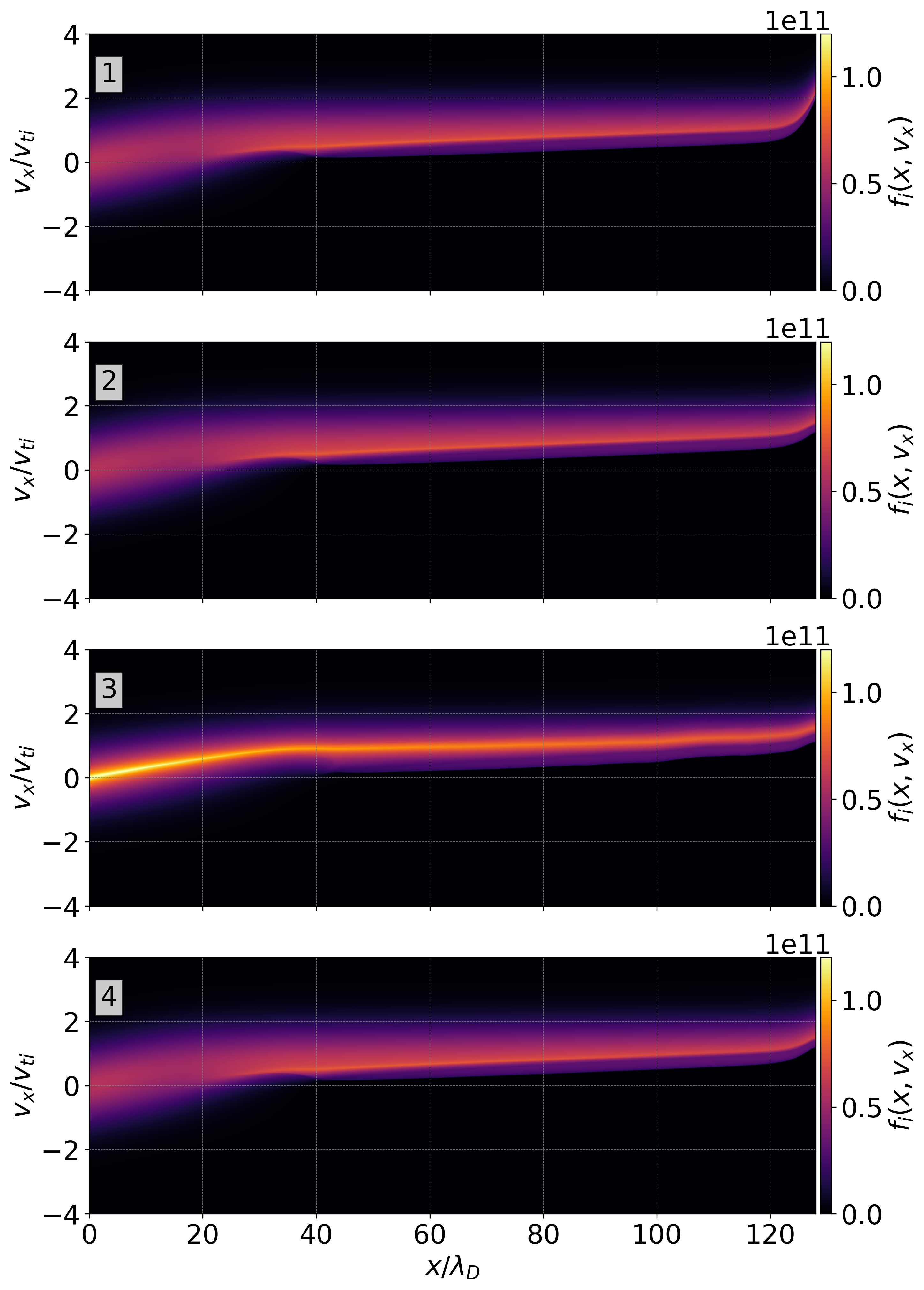}
    \caption{\label{fig:ion_dist} Ion distribution function at $t\omega_{pe} = 8000$ for clean lithium with strong collisions (1), oxidized lithium with strong collisions (2), oxidized lithium with weak collisions (3), and oxidized lithium with the elastic emission removed (4). Only the strongly collisional cases effectively thermalize the source region. The classical case accelerates ions at a greater rate than the SCL cases.}
\end{figure}

\begin{figure}
    \includegraphics[width=0.8\linewidth]{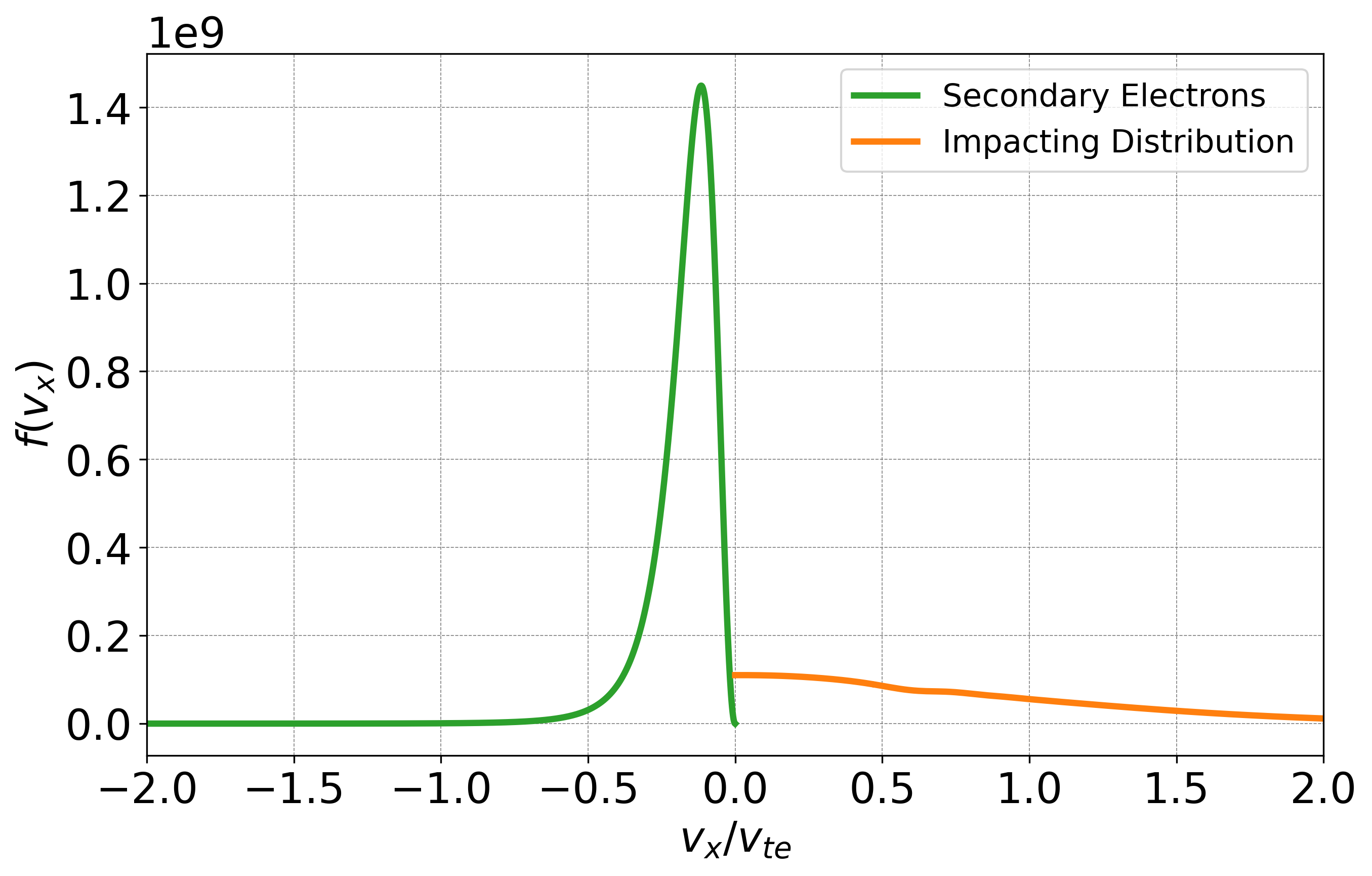}\\
    \includegraphics[width=0.8\linewidth]{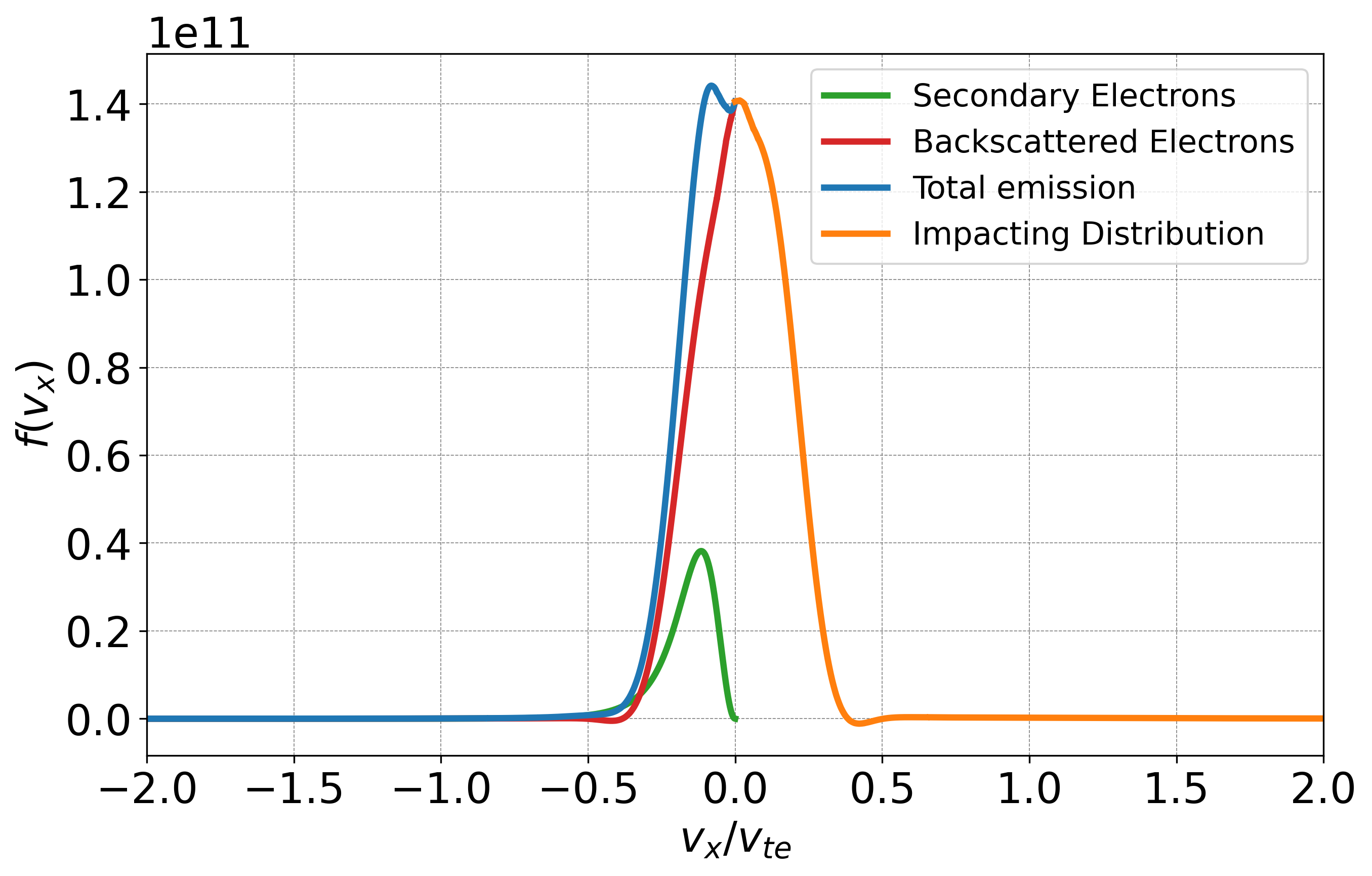}
    \caption{\label{fig:emission_profile} Comparison of emitted distributions of electrons to the impacting electron distribution. In the clean case (top), only true secondary electron emission occurs, so there is only a single population following the Chung-Everhart distribution. For the oxidized case (bottom), the backscattered electrons grow to dominate over the secondary electrons. Note the different magnitudes between the two cases, and the lower average energy in the oxidized case. Low-energy particles are reflected back to the wall by the potential barrier in the SCL sheath, where they re-scatter and repeat the process, leading to the accumulation of low-energy electrons which drives the mean incident energy further towards the backscattering-dominant regime.}
\end{figure}

\section{Results} \label{sec:results}

The cases with the impure lithium experience significantly higher emission, as predicted. Fig.~\ref{fig:sheath_profiles} compares the sheath profiles for each case. The clean lithium (case 1) produces classical sheath profiles for the potential and density. The three oxidized lithium cases produce comparable SCL sheath profiles, with a non-monotonic potential and a sharp accumulation of electrons in the potential dip region. Ion density drops less sharply in the SCL than in the classical case. There is no cold ion trapping in the potential well, as expected. Notably, the case without elastic emission (case 3) produces a significantly lower electron density in the SCL barrier region. As emitted particles are reflected back to the wall at low energy, elastic backscattering is the primary mechanism allowing them to continuously re-emit and accumulate. Thus, this indicates lower rates of particle reflection and re-emission due to the lack of backscattering.\par
Important quantities for the simulation cases are tabulated in Table~\ref{tb:simulation_results}. Despite the high theoretical SEE yields predicted by the lithium data, $\delta_{tot}$ is rapidly driven to less than unity at the wall as the theory expects. Measured at the potential minimum, however, the flux ratio coefficient $\gamma$ is significantly greater than unity for the oxidized cases. For the classical case, these two quantities are of course the same, and significantly less than unity. As these quantities are distinguished by whether or not the electrons reflected by the potential barrier are counted in the flux ratio, a large discrepancy between $\delta$ and $\gamma$ suggests that the reflected electrons are the dominant contributor to emission at the surface. This is born out by the oxidized lithium case where the elastic contribution was dropped (case 4), and $\gamma$ is significantly lower as a result.\par
Fig.~\ref{fig:electron_dist} compares the electron distribution functions of the three cases. All four have a clear beam of emitted electrons proceeding from the surface, with the oxidized beams having a far greater density than in the clean lithium case. Additionally, the oxidized cases show this counterstreaming beam exciting a streaming instability as it is accelerated into the presheath. In cases 2 and 4, this mode reaches the left boundary and is damped by the collisions there. In the marginally collisional case, these modes propagate back to the sheath region without being fully damped, and heavy turbulence is excited in the collisionless region. Fig.~\ref{fig:ion_dist} shows the ion distributions. The marginally collisional case does not succeed in fully thermalizing the source ions, but the three highly collisional cases do. The clean lithium shows more acceleration occuring in the sheath region, as the potential drop is greater for the classical sheath profile than in the SCL sheath.\par
The mechanism by which $\delta_{tot}$ is driven to less than unity at the surface is clear by examining Fig.~\ref{fig:emission_profile}, which decomposes the component distributions of the impacting and emitted electrons late in time. In the oxidized case, backscattering has come to dominate over the true secondary electrons in the emitted distribution. An analysis of the SEE yield shows that in both of the oxidized lithium cases where elastic emission is present over $70\%$ of the total emission energy flux is contributed by elastic backscattering, with the remaining $30\%$ accounted for by the secondary electrons. This is only possible due to an accumulation of low-energy particles at the wall, as was noted when discussing the sheath density profiles. The SCL sheath potential barrier reflects some portion of emitted low-energy electrons back to the wall. Here, they collide and undergo a high rate of elastic backscattering, causing most of them to re-emit. This process repeats, further skewing the distribution towards lower energy and biasing it towards the elastic backscattering regime. This leads to a distribution that is primarily in the backscattering ($\delta_e\leq 1$ by definition) and low $\delta_{ts}$ region of the yield curve, driving the total yield below unity. The prominence of elastic emission is also evident in the electron density profiles of Fig.~\ref{fig:sheath_profiles}. Electron density at the wall is highest in the regular oxidized case, and much lower in the oxidized lithium case without elastic backscattering. This is because lower-energy electrons emit in general far less when elastic backscattering is dropped, and consequently the constant re-emission of these low-energy particles that leads to high accumulation does not occur. In the classical sheath case, this skew does not occur and the distribution into the wall remains near to a Maxwellian. The clean lithium has no backscattering, so the emitted distribution follows the Chung-Everhart distribution exactly.\par
Table~\ref{tb:simulation_results} gives the energy flux densities into the surface, which is the third moment of the distribution $Q_s^+ = \frac{1}{2}m_s\int_0^{\infty}v_x^3f_s(v_x)dv^3$. The clean case has higher ion energy fluxes and lower electron energy fluxes into the surface due to the classical profile. However, the increase in electron flux with increased emission is not balanced by the decrease in ion energy flux. While the oxidized cases both have their ion energy flux decrease by about $40-50\%$ compared to the clean cases, the electron flux more than doubles in the marginally collisional case, and more than quadruples in the highly collisional case. Factoring in that the electron energy flux density was higher than the ion energy flux density even in the classical case, the result is a significant increase in the total energy flux density for the oxidized cases that more than outweighs the decreases in ion energy flux, almost doubling in the regular oxidized case. While all oxidized cases experience higher energy fluxes than the clean case, the high levels of electron trapping associated with the turbulence appear to reduce the total energy flux compared to the highly collisional case despite comparable wall potentials, while the lack of elastic emission in case 4 also marginally reduces the energy flux.\par
Ion particle fluxes are also given in Table~\ref{tb:simulation_results}. Consistent with what has been observed in the previous literature,\cite{campanell2017} the ion particle flux remains largely unaffected by emission at the wall, though the turbulence in case 3 results in a noticeably lower particle flux. From this we note that it is only the energy of impacting ions that is being decreased by the emission, not the raw rate at which they impact the surface.\par
The values $\gamma > 1$ for the oxidized cases suggest the potential for a transition to an inverse sheath in the presence of ion-neutral effects.\cite{campanell2016} If this does indeed happen, it is expected that the trend of increasing electron flux and decreasing ion energy flux will occur at even greater scales.\par
The non-lithium material cases (see Table~\ref{tb:simulation_results}) compare predictably to the lithium. Energy density fluxes trend according to the increase in wall potential. Graphite and tungsten both give classical sheath profiles, while oxidized beryllium is space-charge limited and far closer in magnitude to the oxidized lithium cases. Ion particle fluxes remain comparable regardless of emission strength. Overall, in all cases where strong turbulence is not present, any increase in the emission parameters ($\delta$ and $\gamma$) correlates to a corresponding increase in the wall potential and total energy density flux into the surface.

\begin{table*}
\centering
\begin{tabular}{ |c|c|c|c|c|c|c|c|c|c|c|  }
\hline
Material & $\tilde{\phi}_w$ & $\delta_{tot}$ & $\gamma$ & $Q_{tot}^+\,(\SI{}{\watt/\metre^2})$ & $Q^+_e\,(\SI{}{\watt/\metre^2})$ & $Q^+_i\,(\SI{}{\watt/\metre^2})$ & $\Gamma^+_i\,(\SI{}{\metre^{-2}\second^{-1}})$\\
\hline
(1) Pure Lithium & $-2.58$ & $0.45$ & $0.45$ & $\SI{1.59e5}{}$ & $\SI{5.49e4}{}$ & $\SI{1.05e5}{}$ & $\SI{1.32e21}{}$\\
(2) Oxidized Lithium & $-1.18$ & $0.99$ & $11.0$ & $\SI{2.97e5}{}$ & $\SI{2.34e5}{}$ & $\SI{6.30e4}{}$ & $\SI{1.34e21}{}$\\
(3) Oxidized Lithium (Reduced $\tilde{\nu}_0$) & $-1.18$ & $0.98$ & $8.4$ & $\SI{1.77e5}{}$ & $\SI{1.26e5}{}$ & $\SI{5.08e4}{}$ & $\SI{1.23e21}{}$\\
(4) Oxidized Lithium (No $\delta_e$) & $-1.23$ & $0.94$ & $2.5$ & $\SI{2.78e5}{}$ & $\SI{2.14e5}{}$ & $\SI{6.41e4}{}$ & $\SI{1.34e21}{}$\\
\hline
Tungsten & $-2.10$ & $0.63$ & $0.63$ & $\SI{1.74e5}{}$ & $\SI{8.12e4}{}$ & $\SI{9.30e4}{}$ & $\SI{1.32e21}{}$\\
Graphite & $-1.77$ & $0.73$ & $0.73$ & $\SI{1.96e5}{}$ & $\SI{1.13e5}{}$ & $\SI{8.27e4}{}$ & $\SI{1.32e21}{}$\\
Oxidized Beryllium & $-1.16$ & $0.97$ & $5.0$ & $\SI{2.87e5}{}$ & $\SI{2.24e5}{}$ & $\SI{6.32e4}{}$ & $\SI{1.34e21}{}$\\
\hline
\end{tabular}
\caption{Simulation case results. Normalized wall potential ($\tilde{\phi}_w=q_0\phi_w/T_0$), emission coefficients, energy density fluxes, and ion particle fluxes are given.}
\label{tb:simulation_results}
\end{table*}

\section{Conclusion}

The addition of oxidation and impurities to a lithium surface drastically alters its interactions with a plasma, increasing the surface emission significantly and causing the sheath to take on an SCL profile. The result is increased total energy flux into the wall.\par
The plasma regime chosen here was such that the resulting energy fluxes even when enhanced did not exceed $\SI{300}{\kilo\watt\per\metre^2}$. For devices that regularly see energy fluxes in the $\SI{}{\mega\watt\per\metre^2}$ or $\SI{}{\giga\watt\per\metre^2}$ range,\cite{coenen2011melt,kuang2020divertor} a near-doubling of the energy flux due to material choice might be considerably more dire, particularly in situations such as plasma disruption or edge-localized mode (ELM) bursts where large amounts of energy can be rapidly expelled into the scrape-off layer. At these energy levels, brittle destruction and melting of wall material becomes a serious concern.\cite{wurz2002macroscopic}\par
Results are more promising with regards to sputtering of the wall material. Oxidation does not significantly increase the ion particle flux at the surface, while significantly decreasing the ion energy, suggesting a decrease in the amount of ejected wall material and the resultant wall erosion and plasma contamination.\par
Strong secondary electron emission also drives instability growth which propagates into the boundary region of the plasma. While these modes appear to damp out quickly if collisions are significant, there are potential ramifications on turbulence and plasma performance.\cite{xu1998scrape} When turbulence propagates into the sheath, it has a somewhat mitigating effect on the increase of the impacting energy fluxes.\par
While this work gives insight into what is occurring at the wall, some natural extensions of the work require additional considerations. The most obvious omission highlighted by previous discussion was that of ion-neutral effects, which have the potential to trigger an even more dramatic reversal in sheath behavior through the transition to an inverse sheath.\par
The question of how the emission might impact the edge plasma is left open by this work. In these simulations, the source region was fixed and magnetic fields are neglected. In actual fusion devices, the presence of magnetic fields would add grazing angle effects, requiring at least a second velocity space dimension and an angle-dependent treatment of the emission. The angle is easily accounted for in the emission algorithms,\cite{bradshaw2024} but increase in computational cost for each additional dimension required is significant, something exacerbated by the high resolution needed to fully resolve cold distributions of emitted particles at low energy. Previous studies of secondary electron emission relevant to fusion devices have predicted a suppression of the emission by the magnetic field angle near the wall,\cite{igitkhanov2001attenuation,subba2001including} though Capece et al noted that these studies were largely restricted by their assumption of low emission rates.\cite{capece2016} It has also been noted in the literature that $E\times B$ drift can cause heating of the emitted electrons, further stimulating secondary emission by the reflected particles.\cite{campanell2015self} Future work will prioritize the simulation of fully magnetized sheaths to better examine these predictions.\par
To fully gauge possible effects on plasma performance, better inflow mechanisms at the presheath edge which sustain the sheath without enforcing a steady, smooth condition are necessary. Modifying the input constraint to preserve the power transport into the sheath\cite{stangeby2000plasma} while allowing the upstream temperature to respond to sheath dynamics is one such approach which has been used effectively in past sheath simulations,\cite{li2023plasma} and may prove useful to future work. Edge cooling driven by plasma-material interactions is a notable concern for tokamak design.\cite{stangeby2000plasma} While the primary drive of this cooling is impurity radiation and hydrogenic fuel recycling, secondary electron emission can also drive cooling of the edge plasma,\cite{igitkhanov2001attenuation} which warrants a more careful examination in the future.\par
Regardless of these limitations, the increase in emissivity with lithium oxidation justifies further examination into what precautions are necessary to maximize the benefits of employing lithium walls in the confinement of plasmas.

\begin{acknowledgments}
The work presented here was supported by the U.S. Department of Energy ARPA-E BETHE program under Grant No. DE-AR0001263, and by the NSF Collaborative Research: Frameworks: A Software Ecosystem for Plasma Science and Space Weather Applications project under Award Number 2209471.\par
The simulations presented in this article were performed on computational resources managed and supported by Princeton Research Computing, a consortium of groups including the Princeton Institute for Computational Science and Engineering (PICSciE) and the Office of Information Technology’s High Performance Computing Center and Visualization Laboratory at Princeton University.
\end{acknowledgments}

\section*{Data Availability Statement}

All the simulation results presented in this paper were produced by and are reproducible using the open-source \texttt{Gkeyll} software. Information for obtaining, installing, and running \texttt{Gkeyll} may be found on the documentation site.\cite{gkyldocs} The input files for the simulations used to produce the results in this paper may be acquired from the repository at \url{https://github.com/ammarhakim/gkyl-paper-inp/tree/master/2025_PoP_LithiumSheath}.

\nocite{*}
\bibliography{references}% Produces the bibliography via BibTeX.

\begin{thebibliography}{62}%
\makeatletter
\providecommand \@ifxundefined [1]{%
 \@ifx{#1\undefined}
}%
\providecommand \@ifnum [1]{%
 \ifnum #1\expandafter \@firstoftwo
 \else \expandafter \@secondoftwo
 \fi
}%
\providecommand \@ifx [1]{%
 \ifx #1\expandafter \@firstoftwo
 \else \expandafter \@secondoftwo
 \fi
}%
\providecommand \natexlab [1]{#1}%
\providecommand \enquote  [1]{``#1''}%
\providecommand \bibnamefont  [1]{#1}%
\providecommand \bibfnamefont [1]{#1}%
\providecommand \citenamefont [1]{#1}%
\providecommand \href@noop [0]{\@secondoftwo}%
\providecommand \href [0]{\begingroup \@sanitize@url \@href}%
\providecommand \@href[1]{\@@startlink{#1}\@@href}%
\providecommand \@@href[1]{\endgroup#1\@@endlink}%
\providecommand \@sanitize@url [0]{\catcode `\\12\catcode `\$12\catcode `\&12\catcode `\#12\catcode `\^12\catcode `\_12\catcode `\%12\relax}%
\providecommand \@@startlink[1]{}%
\providecommand \@@endlink[0]{}%
\providecommand \url  [0]{\begingroup\@sanitize@url \@url }%
\providecommand \@url [1]{\endgroup\@href {#1}{\urlprefix }}%
\providecommand \urlprefix  [0]{URL }%
\providecommand \Eprint [0]{\href }%
\providecommand \doibase [0]{http://dx.doi.org/}%
\providecommand \selectlanguage [0]{\@gobble}%
\providecommand \bibinfo  [0]{\@secondoftwo}%
\providecommand \bibfield  [0]{\@secondoftwo}%
\providecommand \translation [1]{[#1]}%
\providecommand \BibitemOpen [0]{}%
\providecommand \bibitemStop [0]{}%
\providecommand \bibitemNoStop [0]{.\EOS\space}%
\providecommand \EOS [0]{\spacefactor3000\relax}%
\providecommand \BibitemShut  [1]{\csname bibitem#1\endcsname}%
\let\auto@bib@innerbib\@empty
%</preamble>
\bibitem [{\citenamefont {Artsimovich}(1972)}]{artsimovich1972}%
  \BibitemOpen
  \bibfield  {author} {\bibinfo {author} {\bibfnamefont {L.}~\bibnamefont {Artsimovich}},\ }\bibfield  {title} {\enquote {\bibinfo {title} {Tokamak devices},}\ }\href@noop {} {\bibfield  {journal} {\bibinfo  {journal} {Nuclear Fusion}\ }\textbf {\bibinfo {volume} {12}},\ \bibinfo {pages} {215} (\bibinfo {year} {1972})}\BibitemShut {NoStop}%
\bibitem [{\citenamefont {Goebel}\ and\ \citenamefont {Katz}(2008)}]{goebel2023}%
  \BibitemOpen
  \bibfield  {author} {\bibinfo {author} {\bibfnamefont {D.~M.}\ \bibnamefont {Goebel}}\ and\ \bibinfo {author} {\bibfnamefont {I.}~\bibnamefont {Katz}},\ }\href@noop {} {\emph {\bibinfo {title} {Fundamentals of electric propulsion: ion and {Hall} thrusters}}}\ (\bibinfo  {publisher} {John Wiley \& Sons, Ltd},\ \bibinfo {year} {2008})\BibitemShut {NoStop}%
\bibitem [{\citenamefont {Stangeby}(2000)}]{stangeby2000plasma}%
  \BibitemOpen
  \bibfield  {author} {\bibinfo {author} {\bibfnamefont {P.~C.}\ \bibnamefont {Stangeby}},\ }\href@noop {} {\emph {\bibinfo {title} {The plasma boundary of magnetic fusion devices}}}\ (\bibinfo  {publisher} {CRC Press},\ \bibinfo {year} {2000})\BibitemShut {NoStop}%
\bibitem [{\citenamefont {Pitts}\ \emph {et~al.}(2005)\citenamefont {Pitts}, \citenamefont {Coad}, \citenamefont {Coster}, \citenamefont {Federici}, \citenamefont {Fundamenski}, \citenamefont {Horacek}, \citenamefont {Krieger}, \citenamefont {Kukushkin}, \citenamefont {Likonen}, \citenamefont {Matthews} \emph {et~al.}}]{pitts2005material}%
  \BibitemOpen
  \bibfield  {author} {\bibinfo {author} {\bibfnamefont {R.}~\bibnamefont {Pitts}}, \bibinfo {author} {\bibfnamefont {J.}~\bibnamefont {Coad}}, \bibinfo {author} {\bibfnamefont {D.}~\bibnamefont {Coster}}, \bibinfo {author} {\bibfnamefont {G.}~\bibnamefont {Federici}}, \bibinfo {author} {\bibfnamefont {W.}~\bibnamefont {Fundamenski}}, \bibinfo {author} {\bibfnamefont {J.}~\bibnamefont {Horacek}}, \bibinfo {author} {\bibfnamefont {K.}~\bibnamefont {Krieger}}, \bibinfo {author} {\bibfnamefont {A.}~\bibnamefont {Kukushkin}}, \bibinfo {author} {\bibfnamefont {J.}~\bibnamefont {Likonen}}, \bibinfo {author} {\bibfnamefont {G.}~\bibnamefont {Matthews}},  \emph {et~al.},\ }\bibfield  {title} {\enquote {\bibinfo {title} {Material erosion and migration in tokamaks},}\ }\href@noop {} {\bibfield  {journal} {\bibinfo  {journal} {Plasma physics and controlled fusion}\ }\textbf {\bibinfo {volume} {47}},\ \bibinfo {pages} {B303} (\bibinfo {year} {2005})}\BibitemShut {NoStop}%
\bibitem [{\citenamefont {Robertson}(2013)}]{robertson2013}%
  \BibitemOpen
  \bibfield  {author} {\bibinfo {author} {\bibfnamefont {S.}~\bibnamefont {Robertson}},\ }\bibfield  {title} {\enquote {\bibinfo {title} {Sheaths in laboratory and space plasmas},}\ }\href@noop {} {\bibfield  {journal} {\bibinfo  {journal} {Plasma Physics and Controlled Fusion}\ }\textbf {\bibinfo {volume} {55}},\ \bibinfo {pages} {093001} (\bibinfo {year} {2013})}\BibitemShut {NoStop}%
\bibitem [{\citenamefont {Hobbs}\ and\ \citenamefont {Wesson}(1967)}]{hobbs1967}%
  \BibitemOpen
  \bibfield  {author} {\bibinfo {author} {\bibfnamefont {G.~D.}\ \bibnamefont {Hobbs}}\ and\ \bibinfo {author} {\bibfnamefont {J.~A.}\ \bibnamefont {Wesson}},\ }\bibfield  {title} {\enquote {\bibinfo {title} {Heat flow through a {Langmuir} sheath in the presence of electron emission},}\ }\href {\doibase 10.1088/0032-1028/9/1/410} {\bibfield  {journal} {\bibinfo  {journal} {Plasma Physics}\ }\textbf {\bibinfo {volume} {9}},\ \bibinfo {pages} {85} (\bibinfo {year} {1967})}\BibitemShut {NoStop}%
\bibitem [{\citenamefont {{The JET Team}}(1990)}]{jet1990results}%
  \BibitemOpen
  \bibfield  {author} {\bibinfo {author} {\bibnamefont {{The JET Team}}},\ }\bibfield  {title} {\enquote {\bibinfo {title} {Results of jet operation with beryllium},}\ }\href {\doibase https://doi.org/10.1016/0022-3115(90)90024-H} {\bibfield  {journal} {\bibinfo  {journal} {Journal of Nuclear Materials}\ }\textbf {\bibinfo {volume} {176-177}},\ \bibinfo {pages} {3--13} (\bibinfo {year} {1990})}\BibitemShut {NoStop}%
\bibitem [{\citenamefont {Neu}\ \emph {et~al.}(2003)\citenamefont {Neu}, \citenamefont {Dux}, \citenamefont {Geier}, \citenamefont {Gruber}, \citenamefont {Kallenbach}, \citenamefont {Krieger}, \citenamefont {Maier}, \citenamefont {Pugno}, \citenamefont {Rohde}, \citenamefont {Schweizer} \emph {et~al.}}]{neu2003tungsten}%
  \BibitemOpen
  \bibfield  {author} {\bibinfo {author} {\bibfnamefont {R.}~\bibnamefont {Neu}}, \bibinfo {author} {\bibfnamefont {R.}~\bibnamefont {Dux}}, \bibinfo {author} {\bibfnamefont {A.}~\bibnamefont {Geier}}, \bibinfo {author} {\bibfnamefont {O.}~\bibnamefont {Gruber}}, \bibinfo {author} {\bibfnamefont {A.}~\bibnamefont {Kallenbach}}, \bibinfo {author} {\bibfnamefont {K.}~\bibnamefont {Krieger}}, \bibinfo {author} {\bibfnamefont {H.}~\bibnamefont {Maier}}, \bibinfo {author} {\bibfnamefont {R.}~\bibnamefont {Pugno}}, \bibinfo {author} {\bibfnamefont {V.}~\bibnamefont {Rohde}}, \bibinfo {author} {\bibfnamefont {S.}~\bibnamefont {Schweizer}},  \emph {et~al.},\ }\bibfield  {title} {\enquote {\bibinfo {title} {Tungsten as plasma-facing material in {ASDEX} upgrade},}\ }\href@noop {} {\bibfield  {journal} {\bibinfo  {journal} {Fusion engineering and design}\ }\textbf {\bibinfo {volume} {65}},\ \bibinfo {pages} {367--374} (\bibinfo {year} {2003})}\BibitemShut {NoStop}%
\bibitem [{\citenamefont {Coenen}\ \emph {et~al.}(2011)\citenamefont {Coenen}, \citenamefont {Philipps}, \citenamefont {Brezinsek}, \citenamefont {Pintsuk}, \citenamefont {Uytdenhouwen}, \citenamefont {Wirtz}, \citenamefont {Kreter}, \citenamefont {Sugiyama}, \citenamefont {Kurishita}, \citenamefont {Torikai} \emph {et~al.}}]{coenen2011melt}%
  \BibitemOpen
  \bibfield  {author} {\bibinfo {author} {\bibfnamefont {J.}~\bibnamefont {Coenen}}, \bibinfo {author} {\bibfnamefont {V.}~\bibnamefont {Philipps}}, \bibinfo {author} {\bibfnamefont {S.}~\bibnamefont {Brezinsek}}, \bibinfo {author} {\bibfnamefont {G.}~\bibnamefont {Pintsuk}}, \bibinfo {author} {\bibfnamefont {I.}~\bibnamefont {Uytdenhouwen}}, \bibinfo {author} {\bibfnamefont {M.}~\bibnamefont {Wirtz}}, \bibinfo {author} {\bibfnamefont {A.}~\bibnamefont {Kreter}}, \bibinfo {author} {\bibfnamefont {K.}~\bibnamefont {Sugiyama}}, \bibinfo {author} {\bibfnamefont {H.}~\bibnamefont {Kurishita}}, \bibinfo {author} {\bibfnamefont {Y.}~\bibnamefont {Torikai}},  \emph {et~al.},\ }\bibfield  {title} {\enquote {\bibinfo {title} {Melt-layer ejection and material changes of three different tungsten materials under high heat-flux conditions in the tokamak edge plasma of {TEXTOR}},}\ }\href@noop {} {\bibfield  {journal} {\bibinfo  {journal} {Nuclear fusion}\ }\textbf {\bibinfo {volume} {51}},\ \bibinfo {pages} {113020}
  (\bibinfo {year} {2011})}\BibitemShut {NoStop}%
\bibitem [{\citenamefont {Kuang}\ \emph {et~al.}(2020)\citenamefont {Kuang}, \citenamefont {Ballinger}, \citenamefont {Brunner}, \citenamefont {Canik}, \citenamefont {Creely}, \citenamefont {Gray}, \citenamefont {Greenwald}, \citenamefont {Hughes}, \citenamefont {Irby}, \citenamefont {LaBombard} \emph {et~al.}}]{kuang2020divertor}%
  \BibitemOpen
  \bibfield  {author} {\bibinfo {author} {\bibfnamefont {A.}~\bibnamefont {Kuang}}, \bibinfo {author} {\bibfnamefont {S.}~\bibnamefont {Ballinger}}, \bibinfo {author} {\bibfnamefont {D.}~\bibnamefont {Brunner}}, \bibinfo {author} {\bibfnamefont {J.}~\bibnamefont {Canik}}, \bibinfo {author} {\bibfnamefont {A.}~\bibnamefont {Creely}}, \bibinfo {author} {\bibfnamefont {T.}~\bibnamefont {Gray}}, \bibinfo {author} {\bibfnamefont {M.}~\bibnamefont {Greenwald}}, \bibinfo {author} {\bibfnamefont {J.}~\bibnamefont {Hughes}}, \bibinfo {author} {\bibfnamefont {J.}~\bibnamefont {Irby}}, \bibinfo {author} {\bibfnamefont {B.}~\bibnamefont {LaBombard}},  \emph {et~al.},\ }\bibfield  {title} {\enquote {\bibinfo {title} {Divertor heat flux challenge and mitigation in {SPARC}},}\ }\href@noop {} {\bibfield  {journal} {\bibinfo  {journal} {Journal of Plasma Physics}\ }\textbf {\bibinfo {volume} {86}},\ \bibinfo {pages} {865860505} (\bibinfo {year} {2020})}\BibitemShut {NoStop}%
\bibitem [{\citenamefont {Menard}\ \emph {et~al.}(2012)\citenamefont {Menard}, \citenamefont {Gerhardt}, \citenamefont {Bell}, \citenamefont {Bialek}, \citenamefont {Brooks}, \citenamefont {Canik}, \citenamefont {Chrzanowski}, \citenamefont {Denault}, \citenamefont {Dudek}, \citenamefont {Gates} \emph {et~al.}}]{menard2012overview}%
  \BibitemOpen
  \bibfield  {author} {\bibinfo {author} {\bibfnamefont {J.}~\bibnamefont {Menard}}, \bibinfo {author} {\bibfnamefont {S.}~\bibnamefont {Gerhardt}}, \bibinfo {author} {\bibfnamefont {M.}~\bibnamefont {Bell}}, \bibinfo {author} {\bibfnamefont {J.}~\bibnamefont {Bialek}}, \bibinfo {author} {\bibfnamefont {A.}~\bibnamefont {Brooks}}, \bibinfo {author} {\bibfnamefont {J.}~\bibnamefont {Canik}}, \bibinfo {author} {\bibfnamefont {J.}~\bibnamefont {Chrzanowski}}, \bibinfo {author} {\bibfnamefont {M.}~\bibnamefont {Denault}}, \bibinfo {author} {\bibfnamefont {L.}~\bibnamefont {Dudek}}, \bibinfo {author} {\bibfnamefont {D.}~\bibnamefont {Gates}},  \emph {et~al.},\ }\bibfield  {title} {\enquote {\bibinfo {title} {Overview of the physics and engineering design of {NSTX} upgrade},}\ }\href@noop {} {\bibfield  {journal} {\bibinfo  {journal} {Nuclear Fusion}\ }\textbf {\bibinfo {volume} {52}},\ \bibinfo {pages} {083015} (\bibinfo {year} {2012})}\BibitemShut {NoStop}%
\bibitem [{\citenamefont {Majeski}\ \emph {et~al.}(2009)\citenamefont {Majeski}, \citenamefont {Berzak}, \citenamefont {Gray}, \citenamefont {Kaita}, \citenamefont {Kozub}, \citenamefont {Levinton}, \citenamefont {Lundberg}, \citenamefont {Manickam}, \citenamefont {Pereverzev}, \citenamefont {Snieckus} \emph {et~al.}}]{majeski2009}%
  \BibitemOpen
  \bibfield  {author} {\bibinfo {author} {\bibfnamefont {R.}~\bibnamefont {Majeski}}, \bibinfo {author} {\bibfnamefont {L.}~\bibnamefont {Berzak}}, \bibinfo {author} {\bibfnamefont {T.}~\bibnamefont {Gray}}, \bibinfo {author} {\bibfnamefont {R.}~\bibnamefont {Kaita}}, \bibinfo {author} {\bibfnamefont {T.}~\bibnamefont {Kozub}}, \bibinfo {author} {\bibfnamefont {F.}~\bibnamefont {Levinton}}, \bibinfo {author} {\bibfnamefont {D.}~\bibnamefont {Lundberg}}, \bibinfo {author} {\bibfnamefont {J.}~\bibnamefont {Manickam}}, \bibinfo {author} {\bibfnamefont {G.}~\bibnamefont {Pereverzev}}, \bibinfo {author} {\bibfnamefont {K.}~\bibnamefont {Snieckus}},  \emph {et~al.},\ }\bibfield  {title} {\enquote {\bibinfo {title} {Performance projections for the lithium tokamak experiment ({LTX})},}\ }\href@noop {} {\bibfield  {journal} {\bibinfo  {journal} {Nuclear Fusion}\ }\textbf {\bibinfo {volume} {49}},\ \bibinfo {pages} {055014} (\bibinfo {year} {2009})}\BibitemShut {NoStop}%
\bibitem [{\citenamefont {Majeski}\ \emph {et~al.}(2010)\citenamefont {Majeski}, \citenamefont {Kugel}, \citenamefont {Kaita}, \citenamefont {Avasarala}, \citenamefont {Bell}, \citenamefont {Bell}, \citenamefont {Berzak}, \citenamefont {Beiersdorfer}, \citenamefont {Gerhardt}, \citenamefont {Granstedt} \emph {et~al.}}]{majeski2010}%
  \BibitemOpen
  \bibfield  {author} {\bibinfo {author} {\bibfnamefont {R.}~\bibnamefont {Majeski}}, \bibinfo {author} {\bibfnamefont {H.}~\bibnamefont {Kugel}}, \bibinfo {author} {\bibfnamefont {R.}~\bibnamefont {Kaita}}, \bibinfo {author} {\bibfnamefont {S.}~\bibnamefont {Avasarala}}, \bibinfo {author} {\bibfnamefont {M.}~\bibnamefont {Bell}}, \bibinfo {author} {\bibfnamefont {R.}~\bibnamefont {Bell}}, \bibinfo {author} {\bibfnamefont {L.}~\bibnamefont {Berzak}}, \bibinfo {author} {\bibfnamefont {P.}~\bibnamefont {Beiersdorfer}}, \bibinfo {author} {\bibfnamefont {S.}~\bibnamefont {Gerhardt}}, \bibinfo {author} {\bibfnamefont {E.}~\bibnamefont {Granstedt}},  \emph {et~al.},\ }\bibfield  {title} {\enquote {\bibinfo {title} {The impact of lithium wall coatings on {NSTX} discharges and the engineering of the lithium tokamak experiment ({LTX})},}\ }\href@noop {} {\bibfield  {journal} {\bibinfo  {journal} {Fusion engineering and design}\ }\textbf {\bibinfo {volume} {85}},\ \bibinfo {pages} {1283--1289} (\bibinfo {year}
  {2010})}\BibitemShut {NoStop}%
\bibitem [{\citenamefont {Kugel}\ \emph {et~al.}(2012)\citenamefont {Kugel}, \citenamefont {Allain}, \citenamefont {Bell}, \citenamefont {Bell}, \citenamefont {Diallo}, \citenamefont {Ellis}, \citenamefont {Gerhardt}, \citenamefont {Heim}, \citenamefont {Jaworski}, \citenamefont {Kaita} \emph {et~al.}}]{kugel2012}%
  \BibitemOpen
  \bibfield  {author} {\bibinfo {author} {\bibfnamefont {H.}~\bibnamefont {Kugel}}, \bibinfo {author} {\bibfnamefont {J.}~\bibnamefont {Allain}}, \bibinfo {author} {\bibfnamefont {M.}~\bibnamefont {Bell}}, \bibinfo {author} {\bibfnamefont {R.}~\bibnamefont {Bell}}, \bibinfo {author} {\bibfnamefont {A.}~\bibnamefont {Diallo}}, \bibinfo {author} {\bibfnamefont {R.}~\bibnamefont {Ellis}}, \bibinfo {author} {\bibfnamefont {S.}~\bibnamefont {Gerhardt}}, \bibinfo {author} {\bibfnamefont {B.}~\bibnamefont {Heim}}, \bibinfo {author} {\bibfnamefont {M.}~\bibnamefont {Jaworski}}, \bibinfo {author} {\bibfnamefont {R.}~\bibnamefont {Kaita}},  \emph {et~al.},\ }\bibfield  {title} {\enquote {\bibinfo {title} {{NSTX} plasma operation with a liquid lithium divertor},}\ }\href@noop {} {\bibfield  {journal} {\bibinfo  {journal} {Fusion Engineering and Design}\ }\textbf {\bibinfo {volume} {87}},\ \bibinfo {pages} {1724--1731} (\bibinfo {year} {2012})}\BibitemShut {NoStop}%
\bibitem [{\citenamefont {De~Castro}\ \emph {et~al.}(2021)\citenamefont {De~Castro}, \citenamefont {Moynihan}, \citenamefont {Stemmley}, \citenamefont {Szott},\ and\ \citenamefont {Ruzic}}]{de2021}%
  \BibitemOpen
  \bibfield  {author} {\bibinfo {author} {\bibfnamefont {A.}~\bibnamefont {De~Castro}}, \bibinfo {author} {\bibfnamefont {C.}~\bibnamefont {Moynihan}}, \bibinfo {author} {\bibfnamefont {S.}~\bibnamefont {Stemmley}}, \bibinfo {author} {\bibfnamefont {M.}~\bibnamefont {Szott}}, \ and\ \bibinfo {author} {\bibfnamefont {D.}~\bibnamefont {Ruzic}},\ }\bibfield  {title} {\enquote {\bibinfo {title} {Lithium, a path to make fusion energy affordable},}\ }\href@noop {} {\bibfield  {journal} {\bibinfo  {journal} {Physics of Plasmas}\ }\textbf {\bibinfo {volume} {28}} (\bibinfo {year} {2021})}\BibitemShut {NoStop}%
\bibitem [{\citenamefont {Zakharov}\ \emph {et~al.}(1997)\citenamefont {Zakharov}, \citenamefont {Gorodetsky}, \citenamefont {Alimov}, \citenamefont {Kanashenko},\ and\ \citenamefont {Markin}}]{zakharov1997}%
  \BibitemOpen
  \bibfield  {author} {\bibinfo {author} {\bibfnamefont {A.}~\bibnamefont {Zakharov}}, \bibinfo {author} {\bibfnamefont {A.}~\bibnamefont {Gorodetsky}}, \bibinfo {author} {\bibfnamefont {V.~K.}\ \bibnamefont {Alimov}}, \bibinfo {author} {\bibfnamefont {S.}~\bibnamefont {Kanashenko}}, \ and\ \bibinfo {author} {\bibfnamefont {A.}~\bibnamefont {Markin}},\ }\bibfield  {title} {\enquote {\bibinfo {title} {Hydrogen retention in plasma-facing materials and its consequences on tokamak operation},}\ }\href@noop {} {\bibfield  {journal} {\bibinfo  {journal} {Journal of nuclear materials}\ }\textbf {\bibinfo {volume} {241}},\ \bibinfo {pages} {52--67} (\bibinfo {year} {1997})}\BibitemShut {NoStop}%
\bibitem [{\citenamefont {Bruining}\ and\ \citenamefont {De~Boer}(1938)}]{bruining1938}%
  \BibitemOpen
  \bibfield  {author} {\bibinfo {author} {\bibfnamefont {H.}~\bibnamefont {Bruining}}\ and\ \bibinfo {author} {\bibfnamefont {J.}~\bibnamefont {De~Boer}},\ }\bibfield  {title} {\enquote {\bibinfo {title} {Secondary electron emission: Part i. secondary electron emission of metals},}\ }\href@noop {} {\bibfield  {journal} {\bibinfo  {journal} {Physica}\ }\textbf {\bibinfo {volume} {5}},\ \bibinfo {pages} {17--30} (\bibinfo {year} {1938})}\BibitemShut {NoStop}%
\bibitem [{\citenamefont {Capece}\ \emph {et~al.}(2016)\citenamefont {Capece}, \citenamefont {Patino}, \citenamefont {Raitses},\ and\ \citenamefont {Koel}}]{capece2016}%
  \BibitemOpen
  \bibfield  {author} {\bibinfo {author} {\bibfnamefont {A.}~\bibnamefont {Capece}}, \bibinfo {author} {\bibfnamefont {M.}~\bibnamefont {Patino}}, \bibinfo {author} {\bibfnamefont {Y.}~\bibnamefont {Raitses}}, \ and\ \bibinfo {author} {\bibfnamefont {B.}~\bibnamefont {Koel}},\ }\bibfield  {title} {\enquote {\bibinfo {title} {Secondary electron emission from lithium and lithium compounds},}\ }\href@noop {} {\bibfield  {journal} {\bibinfo  {journal} {Applied Physics Letters}\ }\textbf {\bibinfo {volume} {109}} (\bibinfo {year} {2016})}\BibitemShut {NoStop}%
\bibitem [{\citenamefont {Skinner}\ \emph {et~al.}(2013)\citenamefont {Skinner}, \citenamefont {Sullenberger}, \citenamefont {Koel}, \citenamefont {Jaworski},\ and\ \citenamefont {Kugel}}]{skinner2013plasma}%
  \BibitemOpen
  \bibfield  {author} {\bibinfo {author} {\bibfnamefont {C.}~\bibnamefont {Skinner}}, \bibinfo {author} {\bibfnamefont {R.}~\bibnamefont {Sullenberger}}, \bibinfo {author} {\bibfnamefont {B.~E.}\ \bibnamefont {Koel}}, \bibinfo {author} {\bibfnamefont {M.}~\bibnamefont {Jaworski}}, \ and\ \bibinfo {author} {\bibfnamefont {H.}~\bibnamefont {Kugel}},\ }\bibfield  {title} {\enquote {\bibinfo {title} {Plasma facing surface composition during {NSTX} {Li} experiments},}\ }\href@noop {} {\bibfield  {journal} {\bibinfo  {journal} {Journal of Nuclear Materials}\ }\textbf {\bibinfo {volume} {438}},\ \bibinfo {pages} {S647--S650} (\bibinfo {year} {2013})}\BibitemShut {NoStop}%
\bibitem [{\citenamefont {W{\"u}rz}\ \emph {et~al.}(2002)\citenamefont {W{\"u}rz}, \citenamefont {Bazylev}, \citenamefont {Landman}, \citenamefont {Pestchanyi},\ and\ \citenamefont {Safronov}}]{wurz2002macroscopic}%
  \BibitemOpen
  \bibfield  {author} {\bibinfo {author} {\bibfnamefont {H.}~\bibnamefont {W{\"u}rz}}, \bibinfo {author} {\bibfnamefont {B.}~\bibnamefont {Bazylev}}, \bibinfo {author} {\bibfnamefont {I.}~\bibnamefont {Landman}}, \bibinfo {author} {\bibfnamefont {S.}~\bibnamefont {Pestchanyi}}, \ and\ \bibinfo {author} {\bibfnamefont {V.}~\bibnamefont {Safronov}},\ }\bibfield  {title} {\enquote {\bibinfo {title} {Macroscopic erosion of divertor and first wall armour in future tokamaks},}\ }\href@noop {} {\bibfield  {journal} {\bibinfo  {journal} {Journal of nuclear materials}\ }\textbf {\bibinfo {volume} {307}},\ \bibinfo {pages} {60--68} (\bibinfo {year} {2002})}\BibitemShut {NoStop}%
\bibitem [{\citenamefont {Campanell}(2013)}]{campanell2013}%
  \BibitemOpen
  \bibfield  {author} {\bibinfo {author} {\bibfnamefont {M.~D.}\ \bibnamefont {Campanell}},\ }\bibfield  {title} {\enquote {\bibinfo {title} {Negative plasma potential relative to electron-emitting surfaces},}\ }\href {\doibase 10.1103/PhysRevE.88.033103} {\bibfield  {journal} {\bibinfo  {journal} {Phys. Rev. E}\ }\textbf {\bibinfo {volume} {88}},\ \bibinfo {pages} {033103} (\bibinfo {year} {2013})}\BibitemShut {NoStop}%
\bibitem [{\citenamefont {Campanell}\ and\ \citenamefont {Umansky}(2016)}]{campanell2016}%
  \BibitemOpen
  \bibfield  {author} {\bibinfo {author} {\bibfnamefont {M.~D.}\ \bibnamefont {Campanell}}\ and\ \bibinfo {author} {\bibfnamefont {M.~V.}\ \bibnamefont {Umansky}},\ }\bibfield  {title} {\enquote {\bibinfo {title} {Strongly emitting surfaces unable to float below plasma potential},}\ }\href {\doibase 10.1103/PhysRevLett.116.085003} {\bibfield  {journal} {\bibinfo  {journal} {Phys. Rev. Lett.}\ }\textbf {\bibinfo {volume} {116}},\ \bibinfo {pages} {085003} (\bibinfo {year} {2016})}\BibitemShut {NoStop}%
\bibitem [{\citenamefont {Campanell}\ and\ \citenamefont {Umansky}(2017)}]{campanell2017}%
  \BibitemOpen
  \bibfield  {author} {\bibinfo {author} {\bibfnamefont {M.~D.}\ \bibnamefont {Campanell}}\ and\ \bibinfo {author} {\bibfnamefont {M.}~\bibnamefont {Umansky}},\ }\bibfield  {title} {\enquote {\bibinfo {title} {Are two plasma equilibrium states possible when the emission coefficient exceeds unity?}}\ }\href@noop {} {\bibfield  {journal} {\bibinfo  {journal} {Physics of Plasmas}\ }\textbf {\bibinfo {volume} {24}} (\bibinfo {year} {2017})}\BibitemShut {NoStop}%
\bibitem [{\citenamefont {Cagas}\ \emph {et~al.}(2017)\citenamefont {Cagas}, \citenamefont {Hakim}, \citenamefont {Juno},\ and\ \citenamefont {Srinivasan}}]{cagas2017_sheath}%
  \BibitemOpen
  \bibfield  {author} {\bibinfo {author} {\bibfnamefont {P.}~\bibnamefont {Cagas}}, \bibinfo {author} {\bibfnamefont {A.}~\bibnamefont {Hakim}}, \bibinfo {author} {\bibfnamefont {J.}~\bibnamefont {Juno}}, \ and\ \bibinfo {author} {\bibfnamefont {B.}~\bibnamefont {Srinivasan}},\ }\bibfield  {title} {\enquote {\bibinfo {title} {Continuum kinetic and multi-fluid simulations of classical sheaths},}\ }\href {\doibase 10.1063/1.4976544} {\bibfield  {journal} {\bibinfo  {journal} {Physics of Plasmas}\ }\textbf {\bibinfo {volume} {24}},\ \bibinfo {pages} {022118} (\bibinfo {year} {2017})},\ \Eprint {http://arxiv.org/abs/https://doi.org/10.1063/1.4976544} {https://doi.org/10.1063/1.4976544} \BibitemShut {NoStop}%
\bibitem [{\citenamefont {Skolar}\ \emph {et~al.}(2023)\citenamefont {Skolar}, \citenamefont {Bradshaw}, \citenamefont {Juno},\ and\ \citenamefont {Srinivasan}}]{skolar2023}%
  \BibitemOpen
  \bibfield  {author} {\bibinfo {author} {\bibfnamefont {C.~R.}\ \bibnamefont {Skolar}}, \bibinfo {author} {\bibfnamefont {K.}~\bibnamefont {Bradshaw}}, \bibinfo {author} {\bibfnamefont {J.}~\bibnamefont {Juno}}, \ and\ \bibinfo {author} {\bibfnamefont {B.}~\bibnamefont {Srinivasan}},\ }\bibfield  {title} {\enquote {\bibinfo {title} {Continuum kinetic investigation of the impact of bias potentials in the current saturation regime on sheath formation},}\ }\href@noop {} {\bibfield  {journal} {\bibinfo  {journal} {Physics of Plasmas}\ }\textbf {\bibinfo {volume} {30}} (\bibinfo {year} {2023})}\BibitemShut {NoStop}%
\bibitem [{\citenamefont {Bradshaw}\ and\ \citenamefont {Srinivasan}(2024)}]{bradshaw2024}%
  \BibitemOpen
  \bibfield  {author} {\bibinfo {author} {\bibfnamefont {K.}~\bibnamefont {Bradshaw}}\ and\ \bibinfo {author} {\bibfnamefont {B.}~\bibnamefont {Srinivasan}},\ }\bibfield  {title} {\enquote {\bibinfo {title} {Energy-dependent implementation of secondary electron emission models in continuum kinetic sheath simulations},}\ }\href {\doibase 10.1088/1361-6595/ad331c} {\bibfield  {journal} {\bibinfo  {journal} {Plasma Sources Science and Technology}\ }\textbf {\bibinfo {volume} {33}},\ \bibinfo {pages} {035008} (\bibinfo {year} {2024})}\BibitemShut {NoStop}%
\bibitem [{\citenamefont {Gkeyll}(2025)}]{gkyldocs}%
  \BibitemOpen
  \bibfield  {author} {\bibinfo {author} {\bibnamefont {Gkeyll}},\ }\href@noop {} {} (\bibinfo {year} {2025}),\ \bibinfo {note} {\url{https://gkeyll.readthedocs.io}}\BibitemShut {NoStop}%
\bibitem [{\citenamefont {Cockburn}\ and\ \citenamefont {Shu}(2001)}]{cockburn2001}%
  \BibitemOpen
  \bibfield  {author} {\bibinfo {author} {\bibfnamefont {B.}~\bibnamefont {Cockburn}}\ and\ \bibinfo {author} {\bibfnamefont {C.-W.}\ \bibnamefont {Shu}},\ }\bibfield  {title} {\enquote {\bibinfo {title} {Runge–{Kutta} discontinuous {Galerkin} methods for convection-dominated problems},}\ }\href {\doibase 10.1023/A:1012873910884} {\bibfield  {journal} {\bibinfo  {journal} {Journal of Scientific Computing}\ }\textbf {\bibinfo {volume} {16}},\ \bibinfo {pages} {173--261} (\bibinfo {year} {2001})}\BibitemShut {NoStop}%
\bibitem [{\citenamefont {Juno}\ \emph {et~al.}(2018)\citenamefont {Juno}, \citenamefont {Hakim}, \citenamefont {TenBarge}, \citenamefont {Shi},\ and\ \citenamefont {Dorland}}]{juno2018discontinuous}%
  \BibitemOpen
  \bibfield  {author} {\bibinfo {author} {\bibfnamefont {J.}~\bibnamefont {Juno}}, \bibinfo {author} {\bibfnamefont {A.}~\bibnamefont {Hakim}}, \bibinfo {author} {\bibfnamefont {J.}~\bibnamefont {TenBarge}}, \bibinfo {author} {\bibfnamefont {E.}~\bibnamefont {Shi}}, \ and\ \bibinfo {author} {\bibfnamefont {W.}~\bibnamefont {Dorland}},\ }\bibfield  {title} {\enquote {\bibinfo {title} {Discontinuous {Galerkin} algorithms for fully kinetic plasmas},}\ }\href@noop {} {\bibfield  {journal} {\bibinfo  {journal} {Journal of Computational Physics}\ }\textbf {\bibinfo {volume} {353}},\ \bibinfo {pages} {110--147} (\bibinfo {year} {2018})}\BibitemShut {NoStop}%
\bibitem [{\citenamefont {Hakim}\ and\ \citenamefont {Juno}(2020)}]{hakim2020alias}%
  \BibitemOpen
  \bibfield  {author} {\bibinfo {author} {\bibfnamefont {A.}~\bibnamefont {Hakim}}\ and\ \bibinfo {author} {\bibfnamefont {J.}~\bibnamefont {Juno}},\ }\bibfield  {title} {\enquote {\bibinfo {title} {Alias-free, matrix-free, and quadrature-free discontinuous {Galerkin} algorithms for (plasma) kinetic equations},}\ }in\ \href@noop {} {\emph {\bibinfo {booktitle} {SC20: International Conference for High Performance Computing, Networking, Storage and Analysis}}}\ (\bibinfo {organization} {IEEE},\ \bibinfo {year} {2020})\ pp.\ \bibinfo {pages} {1--15}\BibitemShut {NoStop}%
\bibitem [{\citenamefont {Li}\ \emph {et~al.}(2022{\natexlab{a}})\citenamefont {Li}, \citenamefont {Srinivasan}, \citenamefont {Zhang},\ and\ \citenamefont {Tang}}]{li2022bohm}%
  \BibitemOpen
  \bibfield  {author} {\bibinfo {author} {\bibfnamefont {Y.}~\bibnamefont {Li}}, \bibinfo {author} {\bibfnamefont {B.}~\bibnamefont {Srinivasan}}, \bibinfo {author} {\bibfnamefont {Y.}~\bibnamefont {Zhang}}, \ and\ \bibinfo {author} {\bibfnamefont {X.-Z.}\ \bibnamefont {Tang}},\ }\bibfield  {title} {\enquote {\bibinfo {title} {Bohm criterion of plasma sheaths away from asymptotic limits},}\ }\href@noop {} {\bibfield  {journal} {\bibinfo  {journal} {Physical Review Letters}\ }\textbf {\bibinfo {volume} {128}},\ \bibinfo {pages} {085002} (\bibinfo {year} {2022}{\natexlab{a}})}\BibitemShut {NoStop}%
\bibitem [{\citenamefont {Li}\ \emph {et~al.}(2022{\natexlab{b}})\citenamefont {Li}, \citenamefont {Srinivasan}, \citenamefont {Zhang},\ and\ \citenamefont {Tang}}]{li2022transport}%
  \BibitemOpen
  \bibfield  {author} {\bibinfo {author} {\bibfnamefont {Y.}~\bibnamefont {Li}}, \bibinfo {author} {\bibfnamefont {B.}~\bibnamefont {Srinivasan}}, \bibinfo {author} {\bibfnamefont {Y.}~\bibnamefont {Zhang}}, \ and\ \bibinfo {author} {\bibfnamefont {X.-Z.}\ \bibnamefont {Tang}},\ }\bibfield  {title} {\enquote {\bibinfo {title} {Transport physics dependence of {Bohm} speed in presheath--sheath transition},}\ }\href@noop {} {\bibfield  {journal} {\bibinfo  {journal} {Physics of Plasmas}\ }\textbf {\bibinfo {volume} {29}} (\bibinfo {year} {2022}{\natexlab{b}})}\BibitemShut {NoStop}%
\bibitem [{\citenamefont {Procassini}, \citenamefont {Birdsall},\ and\ \citenamefont {Morse}(1990)}]{Procassini_1990}%
  \BibitemOpen
  \bibfield  {author} {\bibinfo {author} {\bibfnamefont {R.~J.}\ \bibnamefont {Procassini}}, \bibinfo {author} {\bibfnamefont {C.~K.}\ \bibnamefont {Birdsall}}, \ and\ \bibinfo {author} {\bibfnamefont {E.~C.}\ \bibnamefont {Morse}},\ }\bibfield  {title} {\enquote {\bibinfo {title} {A fully kinetic, selfconsistent particle simulation model of the collisionless plasma-sheath region},}\ }\href {\doibase 10.1063/1.859229} {\bibfield  {journal} {\bibinfo  {journal} {Physics of Fluids B}\ }\textbf {\bibinfo {volume} {2}} (\bibinfo {year} {1990}),\ 10.1063/1.859229}\BibitemShut {NoStop}%
\bibitem [{\citenamefont {Jensen}(2018)}]{jensen2018}%
  \BibitemOpen
  \bibfield  {author} {\bibinfo {author} {\bibfnamefont {K.}~\bibnamefont {Jensen}},\ }\href@noop {} {\emph {\bibinfo {title} {Introduction to the Physics of Electron Emission}}}\ (\bibinfo  {publisher} {John Wiley \& Sons, Inc},\ \bibinfo {address} {New Jersey},\ \bibinfo {year} {2018})\ pp.\ \bibinfo {pages} {155--161}\BibitemShut {NoStop}%
\bibitem [{\citenamefont {Taccogna}\ and\ \citenamefont {Mizzi}(2014)}]{taccogna2014dust}%
  \BibitemOpen
  \bibfield  {author} {\bibinfo {author} {\bibfnamefont {F.}~\bibnamefont {Taccogna}}\ and\ \bibinfo {author} {\bibfnamefont {G.}~\bibnamefont {Mizzi}},\ }\bibfield  {title} {\enquote {\bibinfo {title} {Dust in plasma ii. effects of secondary electrons: ionization and surface emission},}\ }\href@noop {} {\bibfield  {journal} {\bibinfo  {journal} {Contributions to Plasma Physics}\ }\textbf {\bibinfo {volume} {54}},\ \bibinfo {pages} {877--888} (\bibinfo {year} {2014})}\BibitemShut {NoStop}%
\bibitem [{\citenamefont {Campanell}\ \emph {et~al.}(2015)\citenamefont {Campanell}, \citenamefont {Wang}, \citenamefont {Kaganovich},\ and\ \citenamefont {Khrabrov}}]{campanell2015self}%
  \BibitemOpen
  \bibfield  {author} {\bibinfo {author} {\bibfnamefont {M.}~\bibnamefont {Campanell}}, \bibinfo {author} {\bibfnamefont {H.}~\bibnamefont {Wang}}, \bibinfo {author} {\bibfnamefont {I.}~\bibnamefont {Kaganovich}}, \ and\ \bibinfo {author} {\bibfnamefont {A.}~\bibnamefont {Khrabrov}},\ }\bibfield  {title} {\enquote {\bibinfo {title} {Self-amplification of electrons emitted from surfaces in plasmas with e$\times$ b fields},}\ }\href@noop {} {\bibfield  {journal} {\bibinfo  {journal} {Plasma Sources Science and Technology}\ }\textbf {\bibinfo {volume} {24}},\ \bibinfo {pages} {034010} (\bibinfo {year} {2015})}\BibitemShut {NoStop}%
\bibitem [{\citenamefont {Furman}\ and\ \citenamefont {Pivi}(2002)}]{furman2002}%
  \BibitemOpen
  \bibfield  {author} {\bibinfo {author} {\bibfnamefont {M.~A.}\ \bibnamefont {Furman}}\ and\ \bibinfo {author} {\bibfnamefont {M.~T.~F.}\ \bibnamefont {Pivi}},\ }\bibfield  {title} {\enquote {\bibinfo {title} {Probabilistic model for the simulation of secondary electron emission},}\ }\href {\doibase 10.1103/PhysRevSTAB.5.124404} {\bibfield  {journal} {\bibinfo  {journal} {Phys. Rev. ST Accel. Beams}\ }\textbf {\bibinfo {volume} {5}},\ \bibinfo {pages} {124404} (\bibinfo {year} {2002})}\BibitemShut {NoStop}%
\bibitem [{\citenamefont {Cimino}\ \emph {et~al.}(2004)\citenamefont {Cimino}, \citenamefont {Collins}, \citenamefont {Furman}, \citenamefont {Pivi}, \citenamefont {Ruggiero}, \citenamefont {Rumolo},\ and\ \citenamefont {Zimmermann}}]{cimino2004}%
  \BibitemOpen
  \bibfield  {author} {\bibinfo {author} {\bibfnamefont {R.}~\bibnamefont {Cimino}}, \bibinfo {author} {\bibfnamefont {I.~R.}\ \bibnamefont {Collins}}, \bibinfo {author} {\bibfnamefont {M.~A.}\ \bibnamefont {Furman}}, \bibinfo {author} {\bibfnamefont {M.}~\bibnamefont {Pivi}}, \bibinfo {author} {\bibfnamefont {F.}~\bibnamefont {Ruggiero}}, \bibinfo {author} {\bibfnamefont {G.}~\bibnamefont {Rumolo}}, \ and\ \bibinfo {author} {\bibfnamefont {F.}~\bibnamefont {Zimmermann}},\ }\bibfield  {title} {\enquote {\bibinfo {title} {Can low-energy electrons affect high-energy physics accelerators?}}\ }\href {\doibase 10.1103/PhysRevLett.93.014801} {\bibfield  {journal} {\bibinfo  {journal} {Phys. Rev. Lett.}\ }\textbf {\bibinfo {volume} {93}},\ \bibinfo {pages} {014801} (\bibinfo {year} {2004})}\BibitemShut {NoStop}%
\bibitem [{\citenamefont {Larciprete}\ \emph {et~al.}(2013)\citenamefont {Larciprete}, \citenamefont {Grosso}, \citenamefont {Commisso}, \citenamefont {Flammini},\ and\ \citenamefont {Cimino}}]{larciprete2013}%
  \BibitemOpen
  \bibfield  {author} {\bibinfo {author} {\bibfnamefont {R.}~\bibnamefont {Larciprete}}, \bibinfo {author} {\bibfnamefont {D.~R.}\ \bibnamefont {Grosso}}, \bibinfo {author} {\bibfnamefont {M.}~\bibnamefont {Commisso}}, \bibinfo {author} {\bibfnamefont {R.}~\bibnamefont {Flammini}}, \ and\ \bibinfo {author} {\bibfnamefont {R.}~\bibnamefont {Cimino}},\ }\bibfield  {title} {\enquote {\bibinfo {title} {Secondary electron yield of {Cu} technical surfaces: Dependence on electron irradiation},}\ }\href {\doibase 10.1103/PhysRevSTAB.16.011002} {\bibfield  {journal} {\bibinfo  {journal} {Phys. Rev. ST Accel. Beams}\ }\textbf {\bibinfo {volume} {16}},\ \bibinfo {pages} {011002} (\bibinfo {year} {2013})}\BibitemShut {NoStop}%
\bibitem [{\citenamefont {Andronov}\ \emph {et~al.}(2013)\citenamefont {Andronov}, \citenamefont {Smirnov}, \citenamefont {Kaganovich}, \citenamefont {Startsev}, \citenamefont {Raitses},\ and\ \citenamefont {Demidov}}]{andronov2013}%
  \BibitemOpen
  \bibfield  {author} {\bibinfo {author} {\bibfnamefont {A.}~\bibnamefont {Andronov}}, \bibinfo {author} {\bibfnamefont {A.}~\bibnamefont {Smirnov}}, \bibinfo {author} {\bibfnamefont {I.}~\bibnamefont {Kaganovich}}, \bibinfo {author} {\bibfnamefont {E.}~\bibnamefont {Startsev}}, \bibinfo {author} {\bibfnamefont {Y.}~\bibnamefont {Raitses}}, \ and\ \bibinfo {author} {\bibfnamefont {V.}~\bibnamefont {Demidov}},\ }\bibfield  {title} {\enquote {\bibinfo {title} {Secondary electron emission yield in the limit of low electron energy},}\ }\href@noop {} {\bibfield  {journal} {\bibinfo  {journal} {arXiv preprint arXiv:1309.4658}\ } (\bibinfo {year} {2013})}\BibitemShut {NoStop}%
\bibitem [{\citenamefont {Cimino}\ \emph {et~al.}(2015)\citenamefont {Cimino}, \citenamefont {Gonzalez}, \citenamefont {Larciprete}, \citenamefont {Di~Gaspare}, \citenamefont {Iadarola},\ and\ \citenamefont {Rumolo}}]{cimino2015}%
  \BibitemOpen
  \bibfield  {author} {\bibinfo {author} {\bibfnamefont {R.}~\bibnamefont {Cimino}}, \bibinfo {author} {\bibfnamefont {L.~A.}\ \bibnamefont {Gonzalez}}, \bibinfo {author} {\bibfnamefont {R.}~\bibnamefont {Larciprete}}, \bibinfo {author} {\bibfnamefont {A.}~\bibnamefont {Di~Gaspare}}, \bibinfo {author} {\bibfnamefont {G.}~\bibnamefont {Iadarola}}, \ and\ \bibinfo {author} {\bibfnamefont {G.}~\bibnamefont {Rumolo}},\ }\bibfield  {title} {\enquote {\bibinfo {title} {Detailed investigation of the low energy secondary electron yield of technical {Cu} and its relevance for the {LHC}},}\ }\href {\doibase 10.1103/PhysRevSTAB.18.051002} {\bibfield  {journal} {\bibinfo  {journal} {Phys. Rev. ST Accel. Beams}\ }\textbf {\bibinfo {volume} {18}},\ \bibinfo {pages} {051002} (\bibinfo {year} {2015})}\BibitemShut {NoStop}%
\bibitem [{\citenamefont {Bronshtein}\ and\ \citenamefont {Roshchin}(1958)}]{bronshtein1958}%
  \BibitemOpen
  \bibfield  {author} {\bibinfo {author} {\bibfnamefont {I.}~\bibnamefont {Bronshtein}}\ and\ \bibinfo {author} {\bibfnamefont {V.}~\bibnamefont {Roshchin}},\ }\bibfield  {title} {\enquote {\bibinfo {title} {Reflection of electrons and secondary electron emission from some metal surfaces at low primary electron energy},}\ }\href@noop {} {\bibfield  {journal} {\bibinfo  {journal} {Sov. J. Tech.-Phys}\ ,\ \bibinfo {pages} {2271}} (\bibinfo {year} {1958})}\BibitemShut {NoStop}%
\bibitem [{\citenamefont {Khan}, \citenamefont {Hobson},\ and\ \citenamefont {Armstrong}(1963)}]{khan1963}%
  \BibitemOpen
  \bibfield  {author} {\bibinfo {author} {\bibfnamefont {I.}~\bibnamefont {Khan}}, \bibinfo {author} {\bibfnamefont {J.}~\bibnamefont {Hobson}}, \ and\ \bibinfo {author} {\bibfnamefont {R.}~\bibnamefont {Armstrong}},\ }\bibfield  {title} {\enquote {\bibinfo {title} {Reflection and diffraction of slow electrons from single crystals of tungsten},}\ }\href@noop {} {\bibfield  {journal} {\bibinfo  {journal} {Physical Review}\ }\textbf {\bibinfo {volume} {129}},\ \bibinfo {pages} {1513} (\bibinfo {year} {1963})}\BibitemShut {NoStop}%
\bibitem [{\citenamefont {Yakubova}\ and\ \citenamefont {Gorbatyi}(1970)}]{yakubova1970}%
  \BibitemOpen
  \bibfield  {author} {\bibinfo {author} {\bibfnamefont {Z.}~\bibnamefont {Yakubova}}\ and\ \bibinfo {author} {\bibfnamefont {N.}~\bibnamefont {Gorbatyi}},\ }\bibfield  {title} {\enquote {\bibinfo {title} {Reflection of slow electrons from the (110) face of a tungsten crystal},}\ }\href@noop {} {\bibfield  {journal} {\bibinfo  {journal} {Soviet Physics Journal}\ }\textbf {\bibinfo {volume} {13}},\ \bibinfo {pages} {1477--1482} (\bibinfo {year} {1970})}\BibitemShut {NoStop}%
\bibitem [{\citenamefont {Cazaux}(2012)}]{cazaux2012}%
  \BibitemOpen
  \bibfield  {author} {\bibinfo {author} {\bibfnamefont {J.}~\bibnamefont {Cazaux}},\ }\bibfield  {title} {\enquote {\bibinfo {title} {Reflectivity of very low energy electrons (< 10 ev) from solid surfaces: Physical and instrumental aspects},}\ }\href@noop {} {\bibfield  {journal} {\bibinfo  {journal} {Journal of Applied Physics}\ }\textbf {\bibinfo {volume} {111}} (\bibinfo {year} {2012})}\BibitemShut {NoStop}%
\bibitem [{\citenamefont {Everhart}(1960)}]{everhart1960}%
  \BibitemOpen
  \bibfield  {author} {\bibinfo {author} {\bibfnamefont {T.}~\bibnamefont {Everhart}},\ }\bibfield  {title} {\enquote {\bibinfo {title} {Simple theory concerning the reflection of electrons from solids},}\ }\href@noop {} {\bibfield  {journal} {\bibinfo  {journal} {Journal of Applied Physics}\ }\textbf {\bibinfo {volume} {31}},\ \bibinfo {pages} {1483--1490} (\bibinfo {year} {1960})}\BibitemShut {NoStop}%
\bibitem [{\citenamefont {Shih}\ \emph {et~al.}(2002)\citenamefont {Shih}, \citenamefont {Yater}, \citenamefont {Hor},\ and\ \citenamefont {Abrams}}]{shih2002secondary}%
  \BibitemOpen
  \bibfield  {author} {\bibinfo {author} {\bibfnamefont {A.}~\bibnamefont {Shih}}, \bibinfo {author} {\bibfnamefont {J.}~\bibnamefont {Yater}}, \bibinfo {author} {\bibfnamefont {C.}~\bibnamefont {Hor}}, \ and\ \bibinfo {author} {\bibfnamefont {R.}~\bibnamefont {Abrams}},\ }\bibfield  {title} {\enquote {\bibinfo {title} {Secondary electron emission properties of oxidized beryllium {CFA} cathodes},}\ }\href@noop {} {\bibfield  {journal} {\bibinfo  {journal} {IEEE transactions on electron devices}\ }\textbf {\bibinfo {volume} {41}},\ \bibinfo {pages} {2448--2454} (\bibinfo {year} {2002})}\BibitemShut {NoStop}%
\bibitem [{\citenamefont {Coomes}(1939)}]{coomes1939total}%
  \BibitemOpen
  \bibfield  {author} {\bibinfo {author} {\bibfnamefont {E.~A.}\ \bibnamefont {Coomes}},\ }\bibfield  {title} {\enquote {\bibinfo {title} {Total secondary electron emission from tungsten and thorium-coated tungsten},}\ }\href@noop {} {\bibfield  {journal} {\bibinfo  {journal} {Physical Review}\ }\textbf {\bibinfo {volume} {55}},\ \bibinfo {pages} {519} (\bibinfo {year} {1939})}\BibitemShut {NoStop}%
\bibitem [{\citenamefont {Cazaux}(2005)}]{cazaux2005secondary}%
  \BibitemOpen
  \bibfield  {author} {\bibinfo {author} {\bibfnamefont {J.}~\bibnamefont {Cazaux}},\ }\bibfield  {title} {\enquote {\bibinfo {title} {Secondary electron emission yield: graphite and some aromatic hydrocarbons},}\ }\href@noop {} {\bibfield  {journal} {\bibinfo  {journal} {Journal of Physics D: Applied Physics}\ }\textbf {\bibinfo {volume} {38}},\ \bibinfo {pages} {2442} (\bibinfo {year} {2005})}\BibitemShut {NoStop}%
\bibitem [{\citenamefont {Vida}\ \emph {et~al.}(2003)\citenamefont {Vida}, \citenamefont {Josepovits}, \citenamefont {Gy{\H{o}}r},\ and\ \citenamefont {Deak}}]{vida2003characterization}%
  \BibitemOpen
  \bibfield  {author} {\bibinfo {author} {\bibfnamefont {G.}~\bibnamefont {Vida}}, \bibinfo {author} {\bibfnamefont {V.}~\bibnamefont {Josepovits}}, \bibinfo {author} {\bibfnamefont {M.}~\bibnamefont {Gy{\H{o}}r}}, \ and\ \bibinfo {author} {\bibfnamefont {P.}~\bibnamefont {Deak}},\ }\bibfield  {title} {\enquote {\bibinfo {title} {Characterization of tungsten surfaces by simultaneous work function and secondary electron emission measurements},}\ }\href@noop {} {\bibfield  {journal} {\bibinfo  {journal} {Microscopy and Microanalysis}\ }\textbf {\bibinfo {volume} {9}},\ \bibinfo {pages} {337--342} (\bibinfo {year} {2003})}\BibitemShut {NoStop}%
\bibitem [{\citenamefont {de~Castro}, \citenamefont {Oyarz{\'a}bal},\ and\ \citenamefont {Tabar{\'e}s}(2023)}]{de2023cold}%
  \BibitemOpen
  \bibfield  {author} {\bibinfo {author} {\bibfnamefont {A.}~\bibnamefont {de~Castro}}, \bibinfo {author} {\bibfnamefont {E.}~\bibnamefont {Oyarz{\'a}bal}}, \ and\ \bibinfo {author} {\bibfnamefont {F.}~\bibnamefont {Tabar{\'e}s}},\ }\bibfield  {title} {\enquote {\bibinfo {title} {Cold plasma studies on the influence of surface microstructured thickness in the secondary electron emission from tungsten coatings},}\ }\href@noop {} {\bibfield  {journal} {\bibinfo  {journal} {Nuclear Materials and Energy}\ }\textbf {\bibinfo {volume} {34}},\ \bibinfo {pages} {101388} (\bibinfo {year} {2023})}\BibitemShut {NoStop}%
\bibitem [{\citenamefont {Bradshaw}(2024)}]{bradshaw2024_diss}%
  \BibitemOpen
  \bibfield  {author} {\bibinfo {author} {\bibfnamefont {K.}~\bibnamefont {Bradshaw}},\ }\emph {\bibinfo {title} {Emitting Wall Boundary Conditions in Continuum Kinetic Simulations: Unlocking the Effects of Energy-Dependent Material Emission on the Plasma Sheath}},\ \href@noop {} {\bibinfo {type} {{PhD} thesis}},\ \bibinfo  {school} {Virginia Polytechnic Institute and State University}, \bibinfo {address} {Blacksburg, VA} (\bibinfo {year} {2024}),\ \bibinfo {note} {available at \url{https://vtechworks.lib.vt.edu/handle/10919/118137}}\BibitemShut {NoStop}%
\bibitem [{\citenamefont {Chung}\ and\ \citenamefont {Everhart}(1974)}]{chung1974}%
  \BibitemOpen
  \bibfield  {author} {\bibinfo {author} {\bibfnamefont {M.}~\bibnamefont {Chung}}\ and\ \bibinfo {author} {\bibfnamefont {T.}~\bibnamefont {Everhart}},\ }\bibfield  {title} {\enquote {\bibinfo {title} {Simple calculation of energy distribution of low‐energy secondary electrons emitted from metals under electron bombardment},}\ }\href {\doibase 10.1063/1.1663306} {\bibfield  {journal} {\bibinfo  {journal} {Journal of Applied Physics}\ }\textbf {\bibinfo {volume} {45}},\ \bibinfo {pages} {707--709} (\bibinfo {year} {1974})}\BibitemShut {NoStop}%
\bibitem [{\citenamefont {Haynes}(2016)}]{haynes2016}%
  \BibitemOpen
  \bibfield  {author} {\bibinfo {author} {\bibfnamefont {W.~M.}\ \bibnamefont {Haynes}},\ }\href@noop {} {\emph {\bibinfo {title} {CRC handbook of chemistry and physics}}}\ (\bibinfo  {publisher} {CRC press},\ \bibinfo {year} {2016})\BibitemShut {NoStop}%
\bibitem [{\citenamefont {Shukla}(2023)}]{shukla2023gk}%
  \BibitemOpen
  \bibfield  {author} {\bibinfo {author} {\bibfnamefont {A.}~\bibnamefont {Shukla}},\ }\href@noop {} {}\bibinfo {howpublished} {private communication} (\bibinfo {year} {2023})\BibitemShut {NoStop}%
\bibitem [{\citenamefont {Davisson}\ and\ \citenamefont {Germer}(1922)}]{davisson1922thermionic}%
  \BibitemOpen
  \bibfield  {author} {\bibinfo {author} {\bibfnamefont {C.}~\bibnamefont {Davisson}}\ and\ \bibinfo {author} {\bibfnamefont {L.}~\bibnamefont {Germer}},\ }\bibfield  {title} {\enquote {\bibinfo {title} {The thermionic work function of tungsten},}\ }\href@noop {} {\bibfield  {journal} {\bibinfo  {journal} {Physical Review}\ }\textbf {\bibinfo {volume} {20}},\ \bibinfo {pages} {300} (\bibinfo {year} {1922})}\BibitemShut {NoStop}%
\bibitem [{\citenamefont {Rut'kov}, \citenamefont {Afanas'~eva},\ and\ \citenamefont {Gall}(2020)}]{rut2020graphene}%
  \BibitemOpen
  \bibfield  {author} {\bibinfo {author} {\bibfnamefont {E.}~\bibnamefont {Rut'kov}}, \bibinfo {author} {\bibfnamefont {E.}~\bibnamefont {Afanas'~eva}}, \ and\ \bibinfo {author} {\bibfnamefont {N.}~\bibnamefont {Gall}},\ }\bibfield  {title} {\enquote {\bibinfo {title} {Graphene and graphite work function depending on layer number on {Re}},}\ }\href@noop {} {\bibfield  {journal} {\bibinfo  {journal} {Diamond and Related Materials}\ }\textbf {\bibinfo {volume} {101}},\ \bibinfo {pages} {107576} (\bibinfo {year} {2020})}\BibitemShut {NoStop}%
\bibitem [{\citenamefont {Jerner}\ and\ \citenamefont {Magee}(1970)}]{jerner1970effect}%
  \BibitemOpen
  \bibfield  {author} {\bibinfo {author} {\bibfnamefont {R.~C.}\ \bibnamefont {Jerner}}\ and\ \bibinfo {author} {\bibfnamefont {C.~B.}\ \bibnamefont {Magee}},\ }\bibfield  {title} {\enquote {\bibinfo {title} {Effect of surface oxidation on the thermionic work function of beryllium},}\ }\href@noop {} {\bibfield  {journal} {\bibinfo  {journal} {Oxidation of Metals}\ }\textbf {\bibinfo {volume} {2}},\ \bibinfo {pages} {1--9} (\bibinfo {year} {1970})}\BibitemShut {NoStop}%
\bibitem [{\citenamefont {Xu}\ and\ \citenamefont {Cohen}(1998)}]{xu1998scrape}%
  \BibitemOpen
  \bibfield  {author} {\bibinfo {author} {\bibfnamefont {X.}~\bibnamefont {Xu}}\ and\ \bibinfo {author} {\bibfnamefont {R.~H.}\ \bibnamefont {Cohen}},\ }\bibfield  {title} {\enquote {\bibinfo {title} {Scrape-off layer turbulence theory and simulations},}\ }\href@noop {} {\bibfield  {journal} {\bibinfo  {journal} {Contributions to Plasma Physics}\ }\textbf {\bibinfo {volume} {38}},\ \bibinfo {pages} {158--170} (\bibinfo {year} {1998})}\BibitemShut {NoStop}%
\bibitem [{\citenamefont {Igitkhanov}\ and\ \citenamefont {Janeschitz}(2001)}]{igitkhanov2001attenuation}%
  \BibitemOpen
  \bibfield  {author} {\bibinfo {author} {\bibfnamefont {Y.}~\bibnamefont {Igitkhanov}}\ and\ \bibinfo {author} {\bibfnamefont {G.}~\bibnamefont {Janeschitz}},\ }\bibfield  {title} {\enquote {\bibinfo {title} {Attenuation of secondary electron emission from divertor plates due to magnetic field inclination},}\ }\href@noop {} {\bibfield  {journal} {\bibinfo  {journal} {Journal of nuclear materials}\ }\textbf {\bibinfo {volume} {290}},\ \bibinfo {pages} {99--103} (\bibinfo {year} {2001})}\BibitemShut {NoStop}%
\bibitem [{\citenamefont {Subba}\ \emph {et~al.}(2001)\citenamefont {Subba}, \citenamefont {Tskhakaya}, \citenamefont {B{\"u}rbaumer}, \citenamefont {Holzm{\"u}ller-Steinacker}, \citenamefont {Schupfer}, \citenamefont {Stanojevic},\ and\ \citenamefont {Kuhn}}]{subba2001including}%
  \BibitemOpen
  \bibfield  {author} {\bibinfo {author} {\bibfnamefont {F.}~\bibnamefont {Subba}}, \bibinfo {author} {\bibfnamefont {D.}~\bibnamefont {Tskhakaya}}, \bibinfo {author} {\bibfnamefont {H.}~\bibnamefont {B{\"u}rbaumer}}, \bibinfo {author} {\bibfnamefont {U.}~\bibnamefont {Holzm{\"u}ller-Steinacker}}, \bibinfo {author} {\bibfnamefont {N.}~\bibnamefont {Schupfer}}, \bibinfo {author} {\bibfnamefont {M.}~\bibnamefont {Stanojevic}}, \ and\ \bibinfo {author} {\bibfnamefont {S.}~\bibnamefont {Kuhn}},\ }\bibfield  {title} {\enquote {\bibinfo {title} {Including the effect of secondary-electron emission at the divertor targets in code modelling},}\ }\href@noop {} {\bibfield  {journal} {\bibinfo  {journal} {Plasma physics and controlled fusion}\ }\textbf {\bibinfo {volume} {44}},\ \bibinfo {pages} {61} (\bibinfo {year} {2001})}\BibitemShut {NoStop}%
\bibitem [{\citenamefont {Li}\ \emph {et~al.}(2023)\citenamefont {Li}, \citenamefont {Srinivasan}, \citenamefont {Zhang},\ and\ \citenamefont {Tang}}]{li2023plasma}%
  \BibitemOpen
  \bibfield  {author} {\bibinfo {author} {\bibfnamefont {Y.}~\bibnamefont {Li}}, \bibinfo {author} {\bibfnamefont {B.}~\bibnamefont {Srinivasan}}, \bibinfo {author} {\bibfnamefont {Y.}~\bibnamefont {Zhang}}, \ and\ \bibinfo {author} {\bibfnamefont {X.-Z.}\ \bibnamefont {Tang}},\ }\bibfield  {title} {\enquote {\bibinfo {title} {The plasma--sheath transition and bohm criterion in a high recycling divertor},}\ }\href@noop {} {\bibfield  {journal} {\bibinfo  {journal} {Physics of Plasmas}\ }\textbf {\bibinfo {volume} {30}} (\bibinfo {year} {2023})}\BibitemShut {NoStop}%
\end{thebibliography}%

\end{document}